 \documentclass[sigconf]{acmart}

\usepackage{subfig}
\usepackage{graphicx}
\usepackage{subcaption}
\newcommand{\theDevice}{SpellRing}

\AtBeginDocument{%
  \providecommand\BibTeX{{%
    \normalfont B\kern-0.5em{\scshape i\kern-0.25em b}\kern-0.8em\TeX}}}

\copyrightyear{2025}
\acmYear{2025}
\setcopyright{acmlicensed}\acmConference[CHI '25]{CHI Conference on Human Factors in Computing Systems}{April 26-May 1, 2025}{Yokohama, Japan}
\acmBooktitle{CHI Conference on Human Factors in Computing Systems (CHI '25), April 26-May 1, 2025, Yokohama, Japan}
\acmDOI{10.1145/3706598.3713721}
\acmISBN{979-8-4007-1394-1/25/04}

\begin{document}

\title{\theDevice: Recognizing Continuous Fingerspelling in American Sign Language using a Ring}

%

\author{Hyunchul Lim, Nam Anh Dang, Dylan Lee, Tianhong Catherine Yu, Jane Lu, Franklin Mingzhe Li$^*$, Yiqi Jin, Yan Ma$^\dagger$, Xiaojun Bi$^\dagger$, François Guimbretière, and Cheng Zhang}
\email{{hl2365,nd433,dl634,ty274,jdl332,yj226,fvg3, chengzhang}@cornell.edu}
\email{mingzhe2@cs.cmu.edu, {yanma1,xiaojun}@cs.stonybrook.edu}

\orcid{0000-0001-8397-3534}
\affiliation{%
  \institution{Cornell University, Ithaca, New York, USA \\ Carnegie Mellon University$^*$, Pittsburgh, Pennsylvania, USA \\Stony Brook University$^\dagger$, Stony Brook, New York, USA}
   \country{ }}

\renewcommand{\shortauthors}{Hyunchul Lim, et al.}

 \begin{abstract}

\textcolor{black}{Fingerspelling is a critical part of American Sign Language (ASL) recognition and has become an accessible optional text entry method for Deaf and Hard of Hearing (DHH) individuals.} In this paper, we introduce \theDevice, a single smart ring worn on the thumb that recognizes words continuously fingerspelled in ASL. \theDevice{} uses active acoustic sensing (via a microphone and speaker) and an inertial measurement unit (IMU) to track handshape and movement, which are processed through a deep learning algorithm using Connectionist Temporal Classification (CTC) loss. We evaluated the system with 20 ASL signers (13 fluent and 7 learners), using the MacKenzie-Soukoref Phrase Set of 1,164 words and 100 phrases. Offline evaluation yielded top-1 and top-5 word recognition accuracies of 82.45\% (±9.67\%) and 92.42\% (±5.70\%), respectively. In real-time, the system achieved a word error rate (WER) of 0.099 (±0.039) on the phrases. Based on these results, we discuss key lessons and design implications for future minimally obtrusive ASL recognition wearables.

\end{abstract}




\begin{teaserfigure}
  \includegraphics[width=1\linewidth]{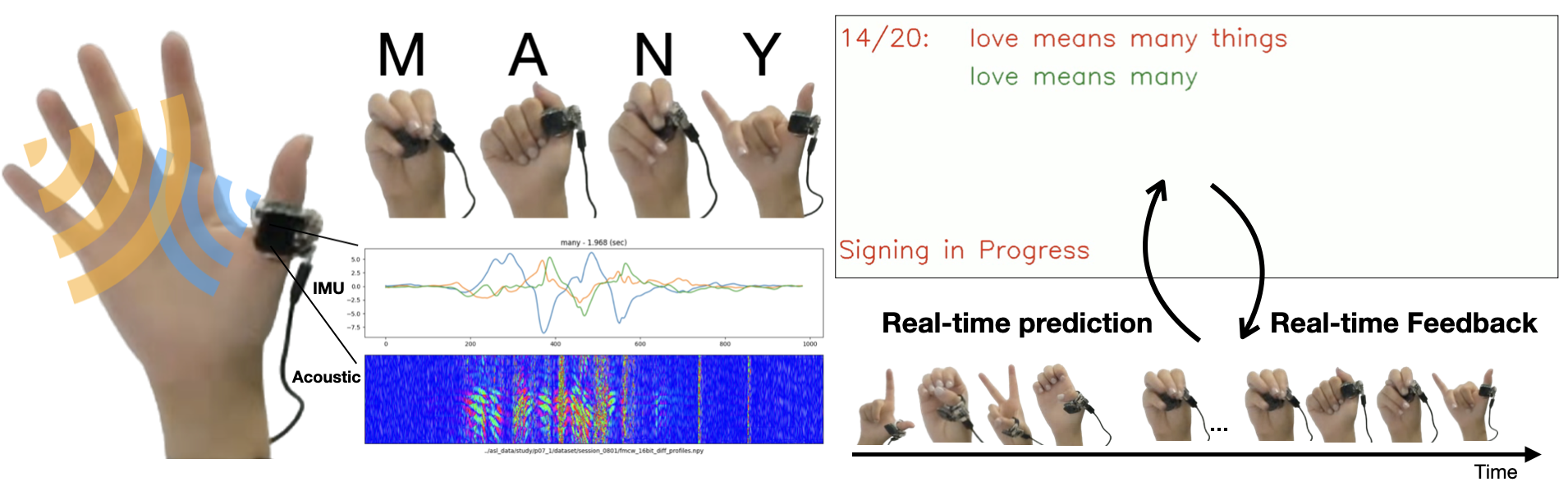}
  \caption{\theDevice{} is a smart ring that uses an AI-powered sensing system designed to recognize continuous fingerspelling}
  \Description{}
  \label{fig:main}
\end{teaserfigure}



\maketitle

\section{INTRODUCTION}

Fingerspelling is used in American Sign Language (ASL) to spell out words without corresponding signs, such as proper nouns, names, and technical terms~\cite{keane2016fingerspelling, hanson1982use}. It is estimated that between 12-35\% of casual ASL conversation comprises fingerspelling~\cite{padden2003alphabet, morere2012fingerspelling}. \textcolor{black}{Recently, fingerspelling has also become a more feasible option for text entry \footnote{https://www.kaggle.com/competitions/asl-fingerspelling} on devices such as mobile phones \cite{hassan2023tap, martin2023fingerspeller}, home assistants \cite{glasser2022analyzing, maria2024alexa}, and VR/AR devices \cite{hirabayashi2019development, gangakhedkar2024fingarspell,fujikawa2019development}, thereby enhancing accessibility for Deaf and Hard of Hearing (DHH) individuals.} Given that ASL fingerspelling involves differentiating handshape, palm orientation, and movement on a single hand, while ASL signs require distinguishing these properties on both hands in addition to sign location (where the sign is articulated relative to the signer's body) and non-manual markers (such as facial expression and body position) \cite{keane2015segmentation, valli1992linguistics}, accurate fingerspelling recognition is a crucial first step in creating comprehensive ASL recognition systems \textcolor{black}{ and is essential for providing DHH individuals with accessible text entry tools.}
While the computer vision community has achieved promising performance in using cameras to detect hands and track finger movements~\cite{aloysius2020understanding, shi2019fingerspelling,hu2020fingertrak}, vision-based approaches are not always portable and raise privacy concerns ( e.g., placing a camera in front of a user's hand is not always feasible.) To address these issues, researchers have explored wearable systems like gloves~\cite{kakoty2018recognition, ahmed2018review, oz2005recognition}, rings~\cite{martin2023fingerspeller,yu2024ring}, and wristbands~\cite{chen2018finger,lee2024echowrist}. However, previous studies suggest that signers prefer for wearable ASL recognition systems to be minimally obtrusive, accurate, and non-disruptive to their natural signing behavior~\cite{kudrinko2022assessing}. Unfortunately, existing wearable ASL systems often fall short in at least one of these areas. Many require bulky hardware~\cite{rizwan2019american, rinalduzzi2021gesture, lee2020sensor, martin2023fingerspeller, singh2023reliable,zhang2018fingerping} that instruments the entire hand or all fingers, which is not practical for everyday use. Furthermore, most systems only recognize isolated manual letters, a significantly easier task for recognition that also forces signers to change their natural signing behavior by pausing between letters while fingerspelling.
\begin{figure}[t]
  \includegraphics[width=0.9\linewidth]{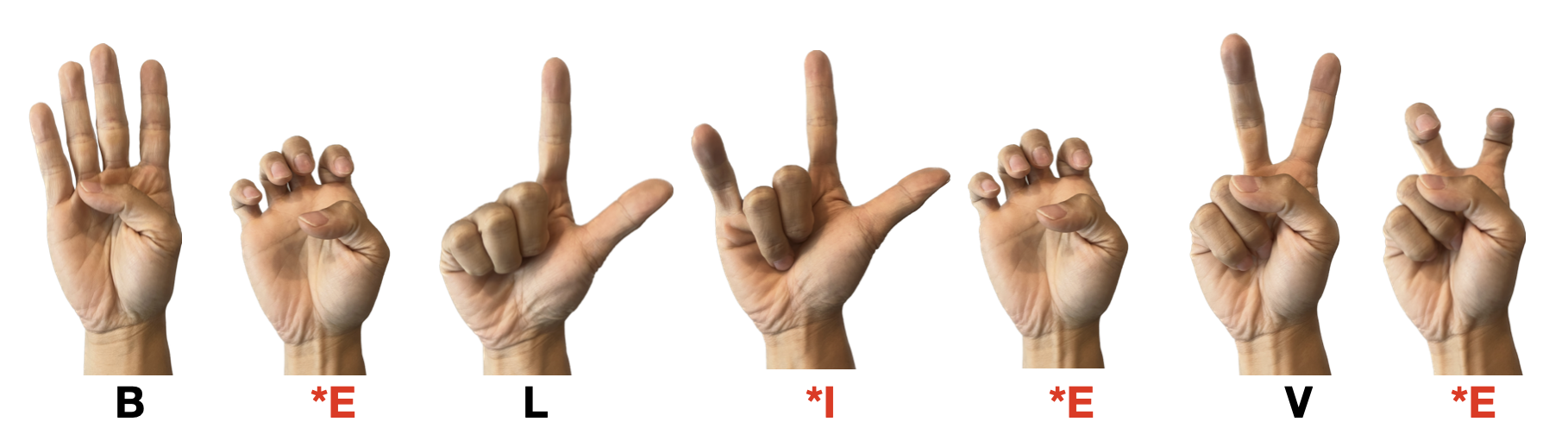}
  \caption{Handshape variation resulting from coarticulation of adjacent fingerspelled letters: Note that the final `E' appears differently from other occurrences of the same letter, and `I' is coarticulated with `L' (in red). }
  \Description{Handshape variation in fingerspelling due to surrounding letters and individual habits: For example, 'E' and 'I' (marked in red) are represented differently.}
  \label{fig:variation}
\end{figure}
While fingerspelling, signers naturally transition between letters continuously, and individual letters are not always distinct. Moreover, the handshape and other properties of a letter may vary depending on articulatory properties of neighboring letters within the word \cite{keane2015segmentation}. As shown in Figure ~\ref{fig:variation}, the same signer may produce different handshapes for the letter ‘E’ within a single word, such as in B-E-L-I-E-V-E; similarly, ‘I' differs from its standard form due to coarticulation with the preceding ‘L'. Accounting for this natural variation complicates continuous fingerspelling recognition. Given these challenges, the central research question we address in this paper is:
\begin{itemize}
    \item Can we design a minimally obtrusive wearable system that can continuously recognize fingerspelled strings without changing signers' behaviors?
\end{itemize}

To tackle this challenge, we introduce \theDevice{}, a deep-learning-powered ring capable of recognizing 1,164 continuously fingerspelled words~\cite{mackenzie2003phrase} without altering signers' fingerspelling behavior. \theDevice{} is a single ring worn on the thumb, utilizing two sensing modalities to capture subtle variations in handshape and movement. The first modality is active acoustic sensing, where a microphone and speaker pair embedded in the ring detect the shape of the entire hand (including all fingers), as demonstrated in~\cite{yu2024ring}. The second is inertial sensing with a gyroscope, which tracks finger motion and helps distinguish between letters with similar handshapes but different palm orientations or movements (e.g., ‘K’ and ‘P’, ‘I’ and ‘J’) \cite{zhang2017fingorbits,zhang2017fingersound}. For instance, ‘K’ and ‘P’ involve identical handshapes but different palm orientations — for ‘K' the palm faces the interlocuter (outward), while for `P' the palm faces downward. The data from these sensors are fused and processed through a custom data processing pipeline and \textcolor{black}{multimodal} deep-learning model incorporating Connectionist Temporal Classification (CTC)~\cite{graves2012connectionist}, which allows the system to recognize English words from a time series of continuous fingerspelling. This algorithm enables the system to recognize words without the need to label individual letters. 

To understand how \theDevice{} performs with signers, we conducted two user studies with 20 participants, focusing on word-level recognition and real-time phrase-level recognition of the system across several days. In both studies, participants naturally fingerspelled words. In Study 1, we evaluated word-level fingerspelling recognition with 9 participants, including 5 ASL learners and 4 fluent signers, collecting approximately 40 hours of data (20,952 words) with an average accuracy of 89.8\%, where recognition for ASL learners (M = 94.38\%, SD = 4.28\%) outperformed that for fluent signers (M = 84.06\%, SD = 9.26\%). In Study 2, we assessed phrase-level fingerspelling recognition with 11 participants in real-time, collecting about 45 hours of training data and testing 100 phrases over approximately 20 hours. The Word Error Rate (WER) was 0.099 (SD = 0.039) with the aid of a language model. After Study 2, a qualitative survey revealed that most participants (N=8) were satisfied with \theDevice{}’s real-time performance. Based on these findings, we provide design recommendations for improving interaction and algorithm development to enhance usability for DHH individuals.

\begin{table*}[t]
\caption{Comparison with Previous Work: \textcolor{black}{ "O" indicates a real-time system, while "X" indicates no real-time capability. Note that SpellRing is a single ring that enables continuous real-time fingerspelling recognition, as evaluated by DHH users.}}
\Description{Comparison with Other Previous work}
\begin{tabular}{r|c|c|c|c|c|c}
\hline
     & \textbf{\textcolor{black}{Evaluation Set}} & \textbf{Form factor} & \textbf{Types} & \textbf{Real-time} & \textbf{Acc.} & \textbf{\# of signers}            \\ \hline
Jani, et al.    \cite{jani2018sensor} & 26 ASL letters   & Data Glove           & isolated       & \textcolor{black}{"O"}                 & 96.5\%       & 8 (N/A)                                    \\ \hline
Yoon, et al. \cite{yoon2012adaptive}   & 24 ASL letters   & 5DT DataGlove        & isolated       & \textcolor{black}{"O"}                       & 91.1\%       & 5 (N/A)                                    \\ \hline
  Savur, et al. \cite{savur2016american}  & 26 ASL letters   & Myo Armband          & isolated       & \textcolor{black}{"O"}                       & 60.8\%       & 10 (all DHH)                                   \\ \hline
 Saquib, et al. \cite{saquib2020application}   & 26 ASL letters   & Data Glove           & isolated       & \textcolor{black}{"O"}                      & 96.0\%         & 5 (N/A)                                \\ \hline
  Lee, et al. \cite{lee2020sensor}   & 27 words         & IMUs on fingertips   & continuous     & \textcolor{black}{"X"}                       & 99.8\%       & 12 (All ASL Learners)                   \\ \hline
  Martin, et al. \cite{martin2023fingerspeller}   & 1164 words       & 5 smart rings        & continuous     & \textcolor{black}{"X"}                   & 91.0\%         & 3 (1 DHH, 2 Experienced)    \\ \hline
SpellRing & 1164 words       & Single ring          & continuous     & \textcolor{black}{"O"}                      & 82.5\%       & 20 (13  DHH, 7 ASL Learners) \\ \hline
\end{tabular}
\end{table*}
\label{table:1}

The contributions of this paper are:
\begin{itemize}
  \item \textcolor{black}{The first wearable-based real-time ASL fingerspelling recognition system that fuses active acoustic sensing with IMU on a ring.}

  \item A multimodal deep-learning pipeline that integrates active acoustic sensing and motion data for continuous fingerspelled word recognition, using Connectionist Temporal Classification (CTC) loss.

  \item A comprehensive evaluation involving 20 ASL signers (13 fluent signers and 7 ASL learners), demonstrating the system's performance across a range of signing experiences, speeds, and personal habits.

  \item A discussion on the opportunities and challenges in designing AI-powered wearables to support DHH individuals in ASL communication.
\end{itemize}

\section{Related Work}
Here, we will discuss prior work that recognizes ASL fingerspelling using cameras, and different types of wearable form factors.

\subsection{Vision-based Approach}
ASL fingerspelling involves complex handshapes that necessitate tracking all fingers for recognition. Thus, early-stage research focused on recognizing hand appearance using vision-based approaches (e.g., cameras), yielding high performance on 26 isolated English/ ASL letters \cite{feris2005recognition, kim2017lexicon,fowley2021sign, goh2006dynamic}. Most previous studies have focused on recognizing isolated manual letters with high performance, achieving over 95\% accuracy \cite{bohavcek2022sign, du2022full, hosain2021hand}. However, given that recognition of isolated manual letters differs significantly from natural, continuous fingerspelling, recent work has shifted toward continuous fingerspelling recognition. Notably, studies using in-the-wild video datasets, such as the FSWild+ dataset \cite{shi2019fingerspelling}, have shown promising results, with performance reaching up to 71.3\%. 
Along with the development of sequential recognition models (HMM \cite{martin2023fingerspeller,goh2006dynamic}, LSTM \cite{shi2019fingerspelling,pannattee2021novel}, CTC clasification \cite{fayyazsanavi2024fingerspelling, gu2022american, graves2006connectionist}), performance on continuous fingerspelling recognition has improved. However, vision-based methods are limited by their costly setup (e.g., by requiring a camera positioned in front of the signer); wearable systems (such as wearing a glove, multiple rings, or a wristband) have proposed more portable alternatives.

\subsection{Wearable-based Approach}
To address the limitations of the vision-based methods mentioned above, wearable-based approaches have attempted to recognize fingerspelling using various form factors, such as data gloves and wristbands.
The data glove, which includes flex sensors \cite{rizwan2019american, saggio2020sign}, stretchable strain sensors \cite{li2018skingest}, magnetic sensors \cite{rinalduzzi2021gesture}, and/ or IMU sensors \cite{lee2020sensor, martin2023fingerspeller, saggio2020sign}, provides highly reliable information on hand joints, detecting both static and dynamic fingerspelling with over 96\% accuracy \cite{lee2020sensor, saggio2020sign, rizwan2019american}. Despite the reliable and high performance, comfort remains a concern among DHH users \cite{kudrinko2022assessing}, and these devices can impair the dexterity of finger movements, thereby affecting ease of use for DHH individuals. 

A wristband using electromyography (EMG) can track finger muscle movements \cite{singh2023reliable}, providing a more usable form factor with a single device worn on the wrist. \textcolor{black}{It has recently demonstrated promising performance in recognizing isolated manual letters, achieving 99.1\% accuracy \cite{singh2023reliable} and 95.36\% accuracy \cite{paudyal2017dyfav}. Additionally, it has shown the ability to recognize various ASL signs \cite{zhang2019myosign, abreu2016evaluating} with accuracies over 92.4\%.} However, continuous fingerspelling recognition is challenging due to the subtle, continuous transitions between letters and the fine-grained movements involved, which may be difficult for EMG sensors to capture accurately and consistently across different wearing sessions, a widely known issue for EMG sesnsors.  Furthermore, the armband with electrodes is dispreferred among DHH users, especially for its poor ease of use and appearance \cite{zhou2023signquery}, and can struggle to deliver reliable performance due to the inherent limitations of EMG sensors \cite{kudrinko2020wearable, ben2023sign} such as calibration, false muscle detection, and surface preparation (including hair removal). \textcolor{black}{Recent work, EchoWrist \cite{lee2024echowrist}, demonstrated continuous finger tracking using a wristband with embedded acoustic sensing, capable of tracking various hand gestures, including isolated ASL digits from 0 to 9. However, while it enables hands-free tracking and classification, its sensor placement is susceptible to obstruction by clothing, such as long sleeves.} Therefore, more effort is required to develop a practical fingerspelling recognition device. A ring-shaped device holds potential for achieving both accuracy and comfort. 

\subsection{Ring-based ASL Approach}
 Similar to the glove, a ring-shaped device places IMU sensors on the fingers, achieving similar performance for continuous fingerspelling. Fingerspeller \cite{martin2023fingerspeller} achieved 91\% accuracy on 1164 unique words and demonstrated the ability to recognize fingerspelled words using four rings, with two rings performing at 81\% accuracy on three signers. However, their work \cite{martin2023fingerspeller} showed that more IMUs are needed on each finger to track the complex movements of fingerspelling. Furthermore, how it would perform on a larger group of native/ fluent signers is unknown. Additionally, \cite{zhou2023signquery} showed that this ring-based form factor scores higher on user experience in ease of use, comfort, and appearance, as compared to GyberGlove\footnote{https://www.cyberglovesystems.com/} and the Myo EMG device\footnote{https://wearables.com/products/myo}. However, current technology still requires multiple rings for recognizing continuous fingerspelling, leaving room for improvement on a design factor.

\textcolor{black}{Recently, Ring-a-Pose \cite{yu2024ring} demonstrated the ability to track 3D handshapes using a single ring with active acoustic sensing, achieving reliable performance on a limited set of hand pose gestures, including the isolated 10 ASL digits (0–9), with a joint error of 14.1 mm.} \textcolor{black}{ Based on research by \cite{shi2018american, shi2019fingerspelling}, which utilizes finger tracking with skeleton information for continuous fingerspelling recognition, Ring-a-Pose appears promising for this task. However, its ability to recognize complex and fast continuous fingerspelling remains uncertain. As hand-tracking error increases, research by \cite{taylor2018real} shows a general downward trend in classification accuracy for 26 isolated letters, highlighting the importance of reliable hand tracking for fingerspelling recognition. However, given that Ring-a-Pose has a 14.1 mm joint error compared to the reliable hand-tracking method used in \cite{shi2018american, shi2019fingerspelling}, its potential still needs further exploration.} Furthermore, Ring-a-Pose \cite{yu2024ring} cannot track palm orientation, which is crucial for distinguishing between manual letters that share an identical handshape and differ solely in palm orientation (e.g., `K' and `P', `G' and `Q', `H' and `U') or movement (e.g., `I' and `J'). Its resolution is also insufficient for the rapid pace of fingerspelling, as fast signers can fingerspell up to eight letters per second \cite{hassan2023tap, hanson1982use, quinto2010rates, keane2016fingerspelling}, while the system tracks only one gesture over two seconds. Fingerspelling comprises 26 letters with similarities across many letters (e.g., `A', `S', `M', `N, and `T'; `C' and `O'; `K' and `P'); fingerspelling recognition is thereby a much more complex task than ASL digit recognition. Fast fingerspelling and signers' individual habits, such as skipping letters or coarticulating neighboring letters \cite{keane2015segmentation}, also lead to variations in handshape for the same letter. Unlike isolated gestures in Ring-a-Pose \cite{yu2024ring}, continuous fingerspelling requires sequential processing to differentiate handshape variations and letter transitions. These inherent challenges necessitate a new approach to continuous fingerspelling recognition.
 
 In this paper, we propose \theDevice{}, which integrates active acoustic sensing \cite{yu2024ring} and an IMU on a thumb-mounted ring to track handshape, hand movement, and palm orientation; we investigate how it can be used to recognize continuously fingerspelled words produced by signers ranging from novice to native/ fluent in ASL proficiency. Table 1 summarizes previous work.

\begin{figure*}[htbp]
    \includegraphics[width=\linewidth]{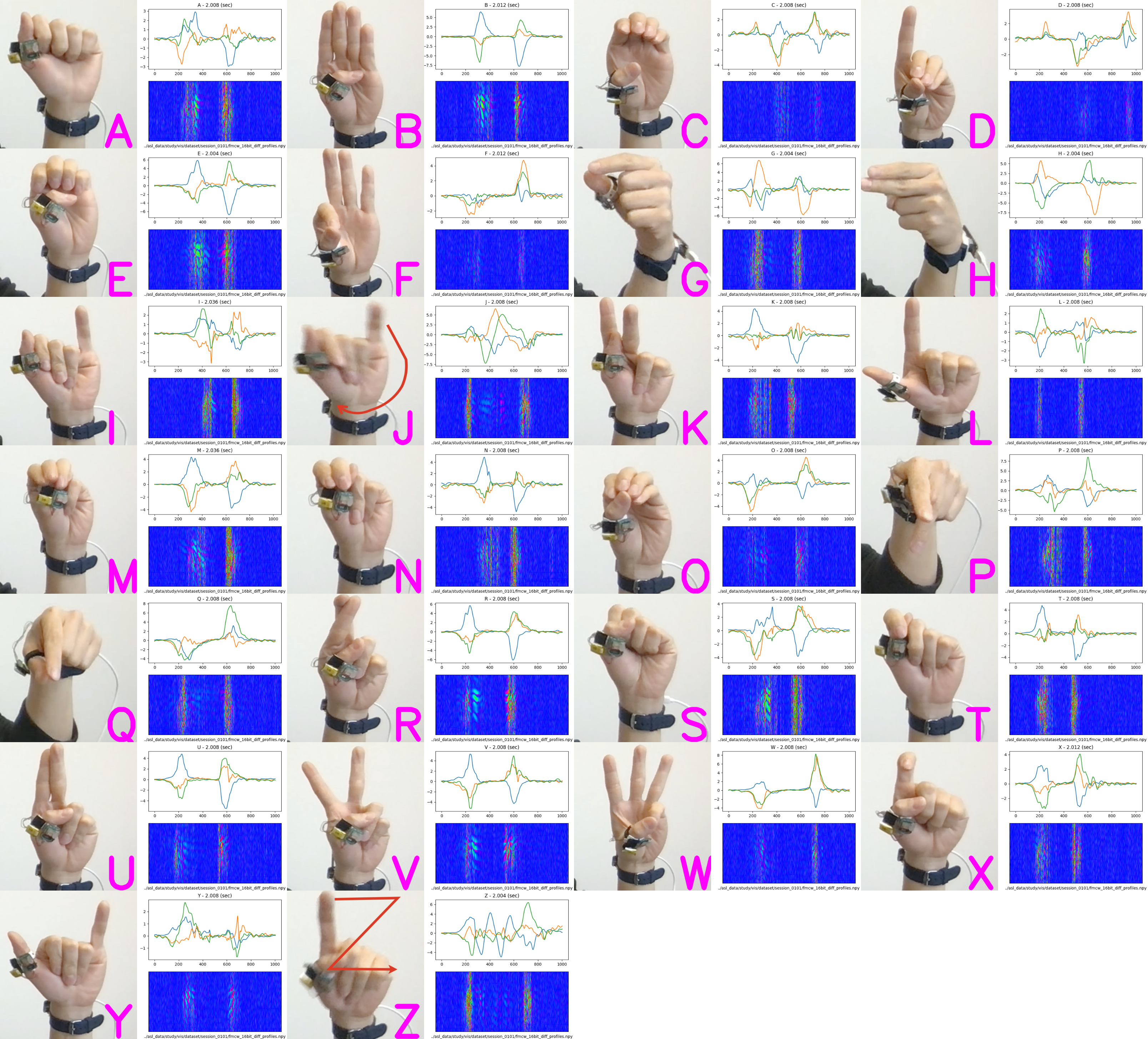}
  \hfill
    \includegraphics[width=\linewidth]{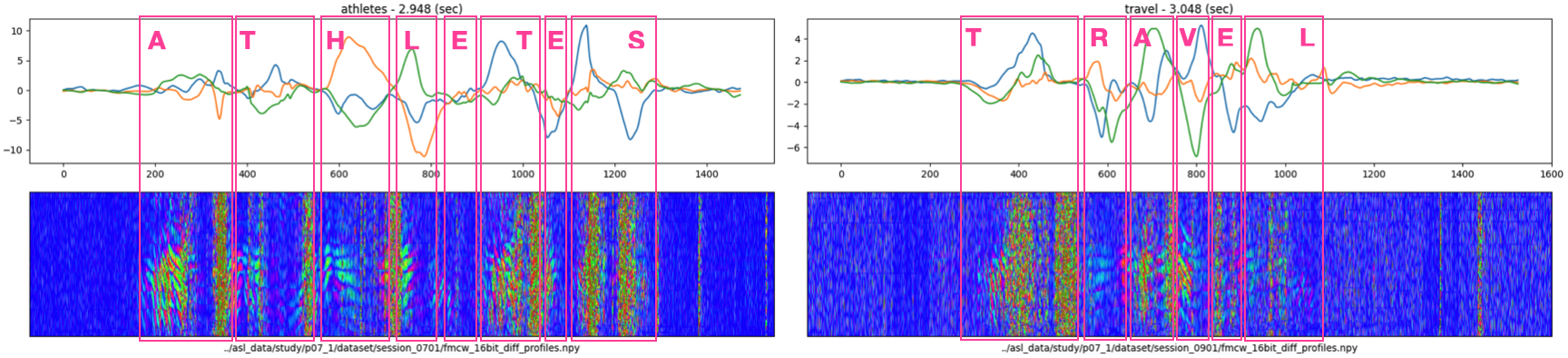}
  \caption{Acoustic and IMU data over 26 isolated English/ASL alphabet letters and continuously fingerspelled words. Continuous fingerspelling adds complexity due to natural flow and quick transitions between letters, which alter sensor values depending on adjacent letters. }
\Description{Acoustic and IMU data over Isolated 26 English manual alphabets and continuous fingerspelled words: Continuous fingerspelling involves added complexity due to the natural flow and quick transitions between letters, which alter sensor values depending on adjacent letters. }
  \label{fig:sensing_principle}
\end{figure*}

\section{SpellRing}

SpellRing is an AI-powered sensing system designed to recognize continuously fingerspelled words using a single ring. This section elaborates on the challenges of designing such a wearable device and details how we developed an AI-powered ring with intelligent sensing methods to achieve accurate recognition.

\subsection{Challenges}
Recognizing continuous fingerspelling poses several unique challenges that make it significantly more complex than recognizing isolated ASL letters:

\subsubsection{Complexity of Handshape, Movement, and Palm Orientation}
American Sign Language (ASL) fingerspelling involves complex combinations of different handshapes, movements, and palm orientations. This poses challenges for accurate fingerspelling recognition. Some letters, such as `A', `S', `M', `N', and `T' (see Figure \ref{fig:sensing_principle}), appear visually similar, while others like `K' and `P', `G' and `Q', or `H' and `U' share the same handshape while differing in palm orientation. Additionally, certain letters (e.g., `J' and `Z') involve specific hand movements, further complicating the recognition process.

\subsubsection{User-Dependent Customized Transitions}
Continuous fingerspelling introduces an additional layer of complexity due to its natural flow and quick transitions between letters (see Figure \ref{fig:sensing_principle}). These transitions vary significantly depending on letter sequences and individual signing behaviors \cite{keane2016fingerspelling, keane2015segmentation}. For instance, the letter `E' can be signed as either a closed or open form, with the open `E' more commonly used during faster fingerspelling, particularly at the beginning or end of a word.
Fluent signers can fingerspell at speeds of 5--8 letters per second \cite{hanson1982use, quinto2010rates, keane2016fingerspelling}, often blending adjacent letters \cite{hassan2023tap, shi2018american, keane2015segmentation}. This high speed increases both efficiency and user-specific signing behaviors, making the accurate recognition of continuous fingerspelling much more challenging than recognizing isolated ASL letters.

\subsubsection{Form Factor vs. Recognition Accuracy Trade-off}
Designing a wearable device for ASL recognition presents a significant challenge in balancing form factor with user experience and recognition accuracy. Glove-based devices with sensors on all fingers can capture detailed poses but are bulky and impractical for daily use, often hindering dexterity. Wristbands, such as EMG sensor bands, offer better usability but struggle with performance issues due to the need for extensive training data across sessions and muscle variability.
Rings with embedded IMUs are more user-friendly, but reliable recognition often requires multiple rings, which can still compromise simplicity. Capturing complex ASL handshapes and movements with a single ring remains a significant challenge, as it must balance unobtrusive design with the ability to capture detailed and reliable data for recognition.

 \begin{table}[t]

\textcolor{black}{
\caption{\textcolor{black}{Performance over Fingerspelling Speed and Sampling Rate. FPS (Frame Per Second), H (Hearing), CODA (Child of Deaf Adults), G (Gender), \# (Max Letters Per Second)}}
\Description{Performance Analysis: Fingerspelling Experience and Sampling Rate}
\begin{tabular}{c|c|c|c|c|c|c}
\hline
   & Experience    & G & Year & \# & FPS - 87     & FPS - 490    \\ \hline
P1 & Leaner, H & M      & 1    & 2                                                                       & 92.99 (2.01) & 93.32 (1.89) \\ \hline
P2 & Leaner, H & F      & 5    & 4                                                                       & 62.67 (3.49) & 86.86 (4.33) \\ \hline
P3 & CODA, H       & M      & 10   & 5                                                                       & 57.93 (3.61) & 86.03 (3.87) \\ \hline
\end{tabular}
}
\end{table}
\label{fig:performance_pilot}
\subsection{Hardware Prototype Design}
To address these challenges, we developed SpellRing, a single-ring system capable of recognizing fingerspelled words at the word level. Our design incorporates two key sensing modalities: acoustic sensing for handshape and IMU sensors for movement.

\begin{figure*}[t]
  \includegraphics[width=0.8\linewidth]{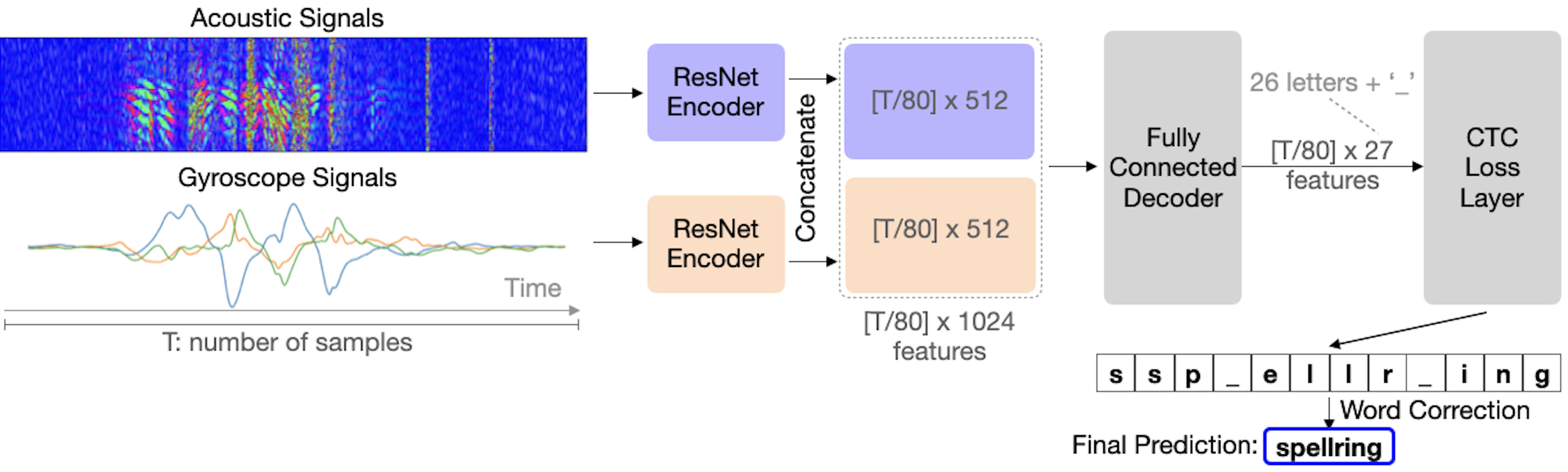}
  \caption{Fusion Model Framework}
  \Description{Fusion Model Framework}
  \label{fig:model}
\end{figure*}

\subsubsection{Single Ring Approach}

SpellRing is designed specifically for the thumb. While Ring-a-Pose \cite{yu2024ring} showed that rings could potentially be placed on all five fingers to track handshape, thumb placement is ideal for ASL recognition. Placing the sensor on other fingers leads to blockage issues, especially when fingerspelling letters such as `A', `S', `M', `N', `L', and `I', and during transitions between these letters. This blockage makes it difficult both to fingerspell and to capture handshape using acoustic sensing. We chose to position the ring on the thumb to minimize these blockage issues. \textcolor{black}{We further discuss the ring's placement and user experience in Section \ref{future}.}

\subsubsection{Sensing Modalities} \textit{1) Active Acoustic Sensing:} For handshape sensing, we chose to adpot active acoustic sensing on the ring. Only requiring low-power and miniature microphone and speakers, this sensing method has shown promising performance in tracking and understanding various body postures on wearables\cite{yu2024ring,li2022eario,li2024eyeecho,li2024gazetrak,mahmud2023posesonic,li2024sonicid,mahmud2024actsonic,mahmud2024munchsonic,sun2023echonose,lee2024echowrist,zhang2023echospeech,mahmud2024wristsonic,zhang2023hpspeech,parikh2024echoguide}. Similar to Ring-a-Pose\cite{yu2024ring}, the ring acts as a `scanner' by emitting inaudible sound waves (frequency range of 20-24 kHz) to scan hand shapes. These sound waves are reflected and refracted by the fingers and received by the microphone on the ring. The preprocessed reflected acoustic signal patterns vary with different hand shapes, leading to precise estimation of handshape \cite{yu2024ring}. \textcolor{black}{However, our earlier experiments (see Table 2) with users of varying fingerspelling skills—especially in speed—using Ring-a-Pose showed that the system struggled to handle rapid fingerspelling of a participant with 10 years of ASL signing experience with a fingerspelling speed of up to 5 letters per second, resulting in an accuracy of 57.93\% on 1,164 words; our experiment is described in detail in Section \ref{experiment}. To capture sufficient information during fast fingerspelling, we increased the sampling rate by six times based on our hardware capabilities. This adjustment reduced the sensing range from 2.06 m (as with Ring-a-Pose \cite{yu2024ring}) to 34.3 cm, focusing more on finger and hand movements and capturing information every 0.12 seconds to classify letters. These changes led to improved performance, achieving an accuracy of 86.03\%---we used this setup for our full experiment.}

\textit{2) IMU for Hand Movement:} To track hand movement and palm orientation, we utilize a gyroscope from the IMU module \cite{zhang2017fingersound,zhang2017fingorbits}. This allows us to measure changes in rotational velocity (angular velocity) around three axes (x, y, and z). By integrating these measurements over time, we can track changes in hand movement, making it easy to distinguish letters with similar handshapes but different palm orientations.
\begin{figure}[b!]
  \includegraphics[width=\linewidth]{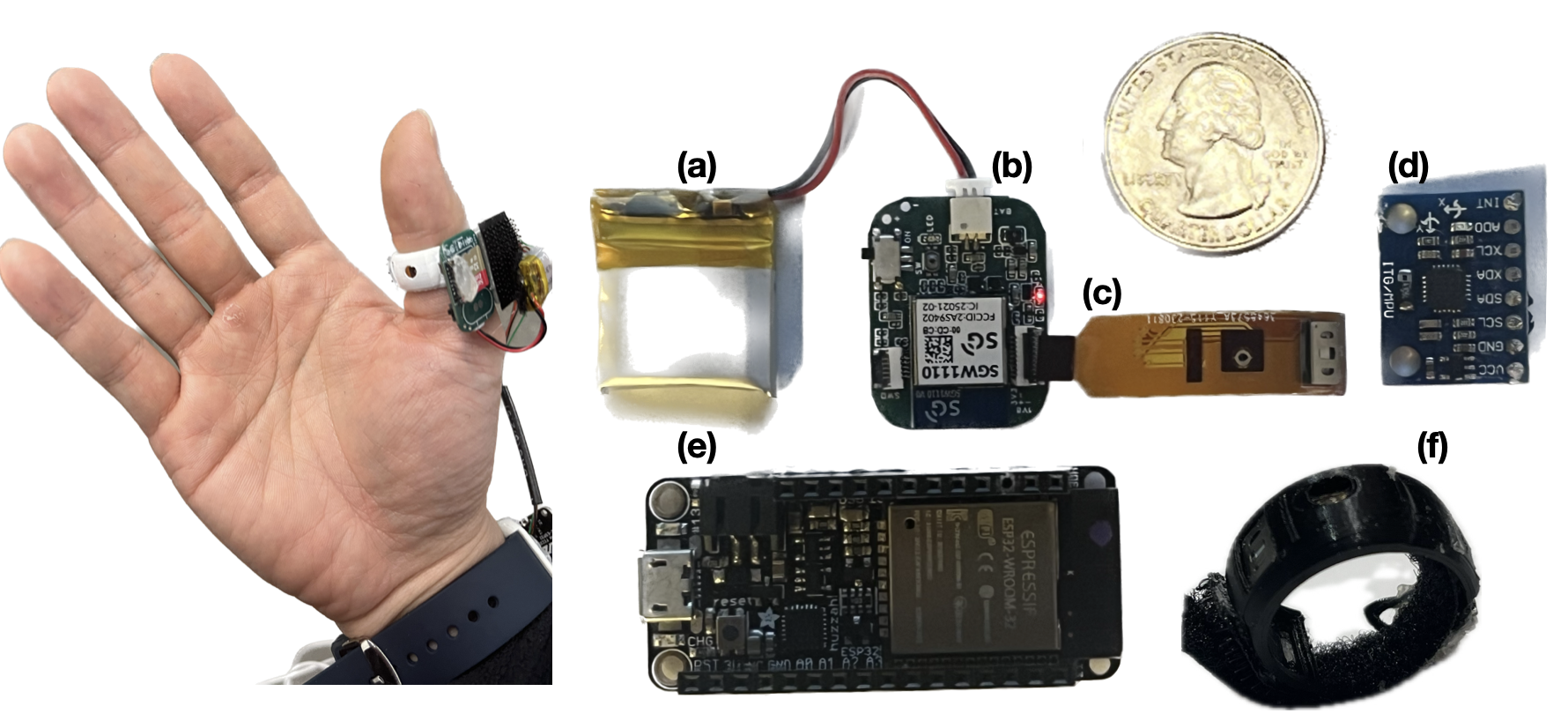}
  \caption{Hardware Prototype: (a) a 3.7V 70mAh LiPo battery, (b) an nRF MCU, (c) a customized Flexible Printed Circuit Board (FPCB) with a microphone and speaker, (d) an IMU sensor board (MPU6050), (e) an ESP32 Feather Board, and (f) a 3D-printed ring case.}
  \Description{Hardware Prototype}
  \label{fig:prototype}
\end{figure}

\subsubsection{Hardware Components}
As shown in Figure \ref{fig:prototype}, the ring incorporates a microphone (TDK-ICS-43434), a speaker (USound UT-P2019), and a customized Flexible Printed Circuit Board (FPCB) \textcolor{black}{(c)} enclosed within a 3D-printed Polylactic Acid (PLA) case \textcolor{black}{(f)}. It also features a microcontroller unit (MCU) \textcolor{black}{(e)}, an SD card for data storage \textcolor{black}{(b)}, and a 3.7V 70mAh LiPo battery \textcolor{black}{(a)}. The ring is powered by the battery, which has a switch for toggling it ON/OFF. Once powered on, the acoustic sensing system initiates and automatically saves data to the SD card until powered off. The IMU sensor board (MPU6050) \textcolor{black}{(d)} includes 6-axis inertial motion sensors (accelerometer and gyroscope), providing three-axis data output at a rate of 150Hz. The IMU is connected to the microcontroller on the wrist via a flexible wire, and the microcontroller transmits the data to an external PC through a flexible USB cable. \textcolor{black}{The acoustic data on the SD card and the IMU data on the PC are then synchronized based on timestamped records.}


\subsection{Algorithms and Data Processing Pipeline}
SpellRing's software pipeline is designed to process multimodal data from the acoustic and IMU sensors and recognize fingerspelled words accurately. Our approach incorporates sophisticated data processing techniques and a deep learning model optimized for sequence recognition.

\subsubsection{Acoustic Data Processing}

Correlation-based frequency modulated continuous wave (C-FMCW) \cite{wang2018c} is used as the transmitted signal for acoustic sensing. The received signals are processed to calculate an echo profile, following methods specified in prior work \cite{yu2024ring,li2022eario,zhang2023echospeech}. These echo profiles encode temporal and spatial information of reflection and diffraction strengths, representing different handshape patterns. To isolate handshape changes from constant environmental reflections, we calculate the difference between consecutive echo frames, generating differential echo profiles. These profiles serve as the input representation of handshape patterns for our deep learning pipeline.

\subsubsection{IMU Data Processing}
Tri-axial gyroscope data, sampled at 150 Hz, is used to track palm orientation and rotational movement. Before feeding them into the deep learning model, we normalize the x, y, and z values and upsample them to synchronize with the acoustic data. This preprocessing step ensures that we can extract synchronized features from our multimodal deep learning pipeline.

\subsubsection{Deep Learning Pipeline}
Our deep learning pipeline leverages Connectionist Temporal Classification (CTC) \cite{graves2006connectionist,zhang2023echospeech}, a method widely employed in sequence labeling tasks, to recognize fingerspelled words continuously without needing to label or segment each letter. Aa shwon in Figure \ref{fig:model}, the model architecture comprises two main components: an acoustic sensing model and an IMU sensing model.


For the acoustic sensing model, \textcolor{black}{we process differential echo profiles using a convolutional neural network (CNN) with ResNet-18 as the backbone. During pooling steps, we apply one-dimensional average pooling along the temporal axis only, rather than both axes, to preserve sequential information.} The IMU sensing model employs a 2D CNN architecture to process IMU data, as our pilot study demonstrated that this approach slightly outperformed a 1D CNN in terms of CTC loss.

The embeddings generated from both modalities are concatenated and then fed into a fully connected dense layer. This is followed by a dropout layer to prevent overfitting, and finally, a softmax function to produce the output probabilities. This multimodal approach allows our system to effectively combine information from both acoustic and motion sensors, enhancing the accuracy of fingerspelling recognition.

\subsubsection{Data Augmentation and Training Scheme}
To enhance performance and streamline training, we adopted several techniques. \textcolor{black}{To enhance the model's adaptability to varying fingerspelling speeds with a fixed window size, we augment the dataset by merging consecutive fingerspelled words, simply concatenating up to four words.} We also apply random noise during training to prevent overfitting and use random padding to handle variable-length inputs. Our training process involves a two-step approach: first training with data from all participants except one, then retraining with the specific participant's data for leave-one-session-out cross-validation. \textcolor{black}{We note that this two-step approach results in a user-dependent model, using 20 sessions collected from each participant over two to three different days, and the following reported results are based on this setup. User-independent results are further discussed in Section \ref{pretrained model}. }

\subsubsection{Word Correction}
To correct potential errors in the model's character sequence predictions, we compute the Levenshtein distance \cite{Levenshtein1965BinaryCC} between the predicted sequence and each unique word in a reference dictionary. The word with the smallest Levenshtein distance was selected as the corrected word, enhancing the overall accuracy of our system. \textcolor{black}{To align our system evaluation with prior literature, specifically for performance comparison with FingerSpeller \cite{martin2023fingerspeller}—such as multi-ring versus single-ring setups—we adopted their evaluation method by using the MacKenzie-Soukoreff phrase set \cite{mackenzie2003phrase} as the reference dictionary.}
\textcolor{black}{However, since the choice of reference dictionary affects the performance of the auto-correction, we discuss its impact using different dictionary sets in Section \ref{auto_correction}.}

\section{User Studies Overview}
To evaluate the effectiveness and usability of \theDevice, we conducted two user studies with signers ranging from novice to native/ fluent in ASL proficiency. The first study focused on word-level recognition, assessing the system's ability to accurately recognize individual fingerspelled words across a diverse group of participants. The second study expanded on these findings by examining phrase-level recognition in real-time scenarios, providing insights into the system's performance in more natural, context-rich environments. These studies aimed to validate SpellRing's performance across different levels of signing experience, explore the impact of signing speed and habits on recognition accuracy, and investigate how users adapt to the system in real-time use. The user study was approved by the Institutional Review Board (IRB) at the authors' institution. Participants were compensated \$40 per hour for their participation in the study.

\section{Word-level Recognition}

\subsection{Purpose and Overview}
The primary objective of this study was to evaluate SpellRing's ability to recognize individual fingerspelled words accurately \textcolor{black}{and compare its performance to that of of multiple-ring setups, such as FingerSpeller \cite{martin2023fingerspeller}.} We aimed to assess the system's performance across a range of users with varying ASL proficiencies, ranging from novice to native. This study also sought to investigate the impact of signing speed and individual habits on recognition accuracy, providing important insights for system optimization and user adaptation strategies.

\subsection{Participants}
We recruited 9 participants (6 male, 3 female, mean age = 23.0, SD = 5.45) to evaluate our system, including 3 Deaf individuals, 1 CODA (Child of Deaf Adults), and 5 hearing ASL learners. The 3 Deaf participants and CODA were fluent signers, using ASL as a primary language. The 5 hearing ASL learners had between 1 and 5 years of ASL learning experience (M = 3.0  SD = 1.4). All participants fingerspelled with their right hand as their dominant hand. Detailed participant information, including proficiency and background, is provided in Table~\ref{fig:table2}.

\subsection{Dataset and Procedure}
\label{experiment}

\begin{figure}[t]
  \includegraphics[width=\linewidth]{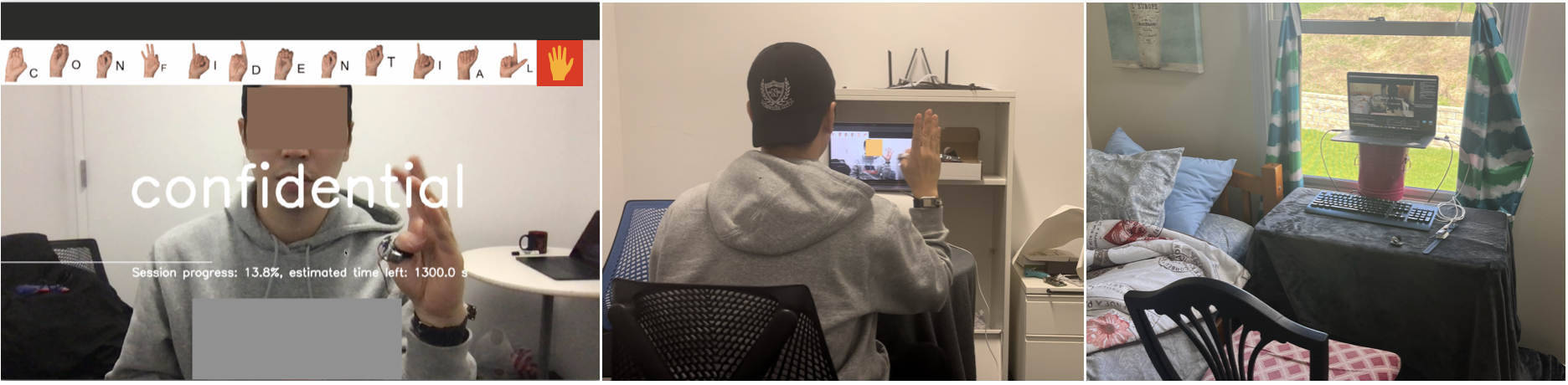}
  \caption{Study Procedures: an English word with guide fingerspelling images (left) and examples of experimental setup locations (e.g., in a study room (center); in a home (right)) }
  \Description{Study Procedures: a word with guide fingerspelling images and the study locations (in a study room and a home)}
  \label{fig:studyprocedure}
\end{figure}

\textcolor{black}{To ensure consistency with FingerSpeller \cite{martin2023fingerspeller}, particularly for performance evaluation, we adopted their dataset and procedure. Since we evaluated our system as an accessible optional text entry method, the MacKenzie-Soukoreff Phrase Set \cite{mackenzie2003phrase} was selected as the standard dataset, which is commonly used to evaluate text entry systems.} This set comprises 500 phrases, totaling 1,164 unique words. The words in this set vary in length, ranging from a minimum of 1 to a maximum of 13, with an average word length of 6.055 (SD: 2.312).

Each participant was tasked with fingerspelling each of these words twice, resulting in a comprehensive dataset of 2328=2*1164 fingerspelled words per participant, which was collected in two rounds of data collection. During each round, a partcipant completed each word once across 10 sessions. Nine of these sessions contained 116 words each, and one session contained 120 words. We carefully ensured that there was no overlap in words between sessions and that each session maintained a similar distribution of letters and word lengths. The average word length across sessions was 6.053 (SD = 0.181), with median values of 5 or 6.

The experiment was conducted in a semi-controlled environment, as illustrated in Figure \ref{fig:studyprocedure}. During each session, participants followed specific guidelines for fingerspelling. They were provided with real-time video feedback of their signing and were allowed to correct any mistakes as needed. To aid ASL learners, we displayed each English word with accompanying guide images of corresponding ASL letters. Participants used their non-dominant hand to press a space key after completing each word, which allowed us to record the start and end times for each fingerspelled sequence to calculate participants' fingerspelling speed. We instructed participants to return to a neutral hand pose between words and to fingerspell at their own comfortable, natural pace. Session durations varied from 5 to 13 minutes, depending on the participant's fingerspelling speed.

Data collection for each participant lasted around 4 hours, split across two to three days for each participant. To assess the effect of device positioning, participants were asked to remove and reattach the ring between sessions. This approach allowed us to collect data under various ring positions, simulating real-world usage scenarios.

\begin{table*}[t]
\caption{Top-N Word Recognition and Participant Information: CODA: Child of Deaf Adults, H: Hearing. Years indicates number of years learning ASL or using it as a primary language.}
\Description{Top-N Word Recognition and Participant Information: CODA (Child of Deaf Adults), H: Hearing. Years indicates number of years learning ASL or using it as a primary language.}
\begin{tabular}{cccc|cccccr}
\hline
\multicolumn{4}{c|}{}                                                                                                  & \multicolumn{6}{c}{Offline Evaulation}                                                                                                                                                                                    \\
\multicolumn{4}{c|}{}                                                                                                  & \multicolumn{6}{c}{\textbf{Word-level}}                                                                                                                                                                                   \\ \hline
\multicolumn{1}{c|}{}     & \multicolumn{1}{c|}{\textbf{Experience}} & \multicolumn{1}{c|}{\textbf{Gender}} & \textbf{Years} & \multicolumn{1}{c|}{\textbf{Top1}} & \multicolumn{1}{c|}{\textbf{Top2}} & \multicolumn{1}{c|}{\textbf{Top3}} & \multicolumn{1}{c|}{\textbf{Top4}} & \multicolumn{1}{c|}{\textbf{Top5}} & \multicolumn{1}{c}{\textbf{LER}} \\ \hline
\multicolumn{1}{c|}{\textcolor{black}{Avg.}} & \multicolumn{1}{c|}{}               & \multicolumn{1}{c|}{}                & \textit{5.44} & \multicolumn{1}{c|}{\textit{89.8}} & \multicolumn{1}{c|}{\textit{92.8}} & \multicolumn{1}{c|}{\textit{94.3}} & \multicolumn{1}{c|}{\textit{95.1}} & \multicolumn{1}{c|}{\textit{95.6}} & \textit{0.131}                   \\ \hline

\multicolumn{1}{c|}{P01}  & \multicolumn{1}{c|}{Learner, H}     & \multicolumn{1}{c|}{F}               & 3             & \multicolumn{1}{c|}{95.338}        & \multicolumn{1}{c|}{97.220}        & \multicolumn{1}{c|}{98.162}        & \multicolumn{1}{c|}{98.501}        & \multicolumn{1}{c|}{98.846}        & 0.068                            \\ \hline
\multicolumn{1}{c|}{P02}  & \multicolumn{1}{c|}{Learner, H}     & \multicolumn{1}{c|}{M}               & 5             & \multicolumn{1}{c|}{89.161}        & \multicolumn{1}{c|}{92.943}        & \multicolumn{1}{c|}{94.796}        & \multicolumn{1}{c|}{95.612}        & \multicolumn{1}{c|}{96.172}        & 0.126                            \\ \hline
\multicolumn{1}{c|}{P03}  & \multicolumn{1}{c|}{Learner, H}     & \multicolumn{1}{c|}{F}               & 3             & \multicolumn{1}{c|}{94.546}        & \multicolumn{1}{c|}{96.565}        & \multicolumn{1}{c|}{97.772}        & \multicolumn{1}{c|}{98.155}        & \multicolumn{1}{c|}{98.371}        & 0.155                            \\ \hline
\multicolumn{1}{c|}{P04}  & \multicolumn{1}{c|}{Learner, H}     & \multicolumn{1}{c|}{M}               & 1             & \multicolumn{1}{c|}{97.851}        & \multicolumn{1}{c|}{98.968}        & \multicolumn{1}{c|}{99.226}        & \multicolumn{1}{c|}{99.527}        & \multicolumn{1}{c|}{99.614}        & 0.042                            \\ \hline
\multicolumn{1}{c|}{P05}  & \multicolumn{1}{c|}{Learner, H}     & \multicolumn{1}{c|}{F}               & 3             & \multicolumn{1}{c|}{95.020}        & \multicolumn{1}{c|}{96.823}        & \multicolumn{1}{c|}{97.599}        & \multicolumn{1}{c|}{98.026}        & \multicolumn{1}{c|}{98.499}        & 0.089                            \\ \hline
\multicolumn{1}{c|}{P06}  & \multicolumn{1}{c|}{Deaf}           & \multicolumn{1}{c|}{M}               & 7             & \multicolumn{1}{c|}{79.840}        & \multicolumn{1}{c|}{83.822}        & \multicolumn{1}{c|}{86.136}        & \multicolumn{1}{c|}{87.358}        & \multicolumn{1}{c|}{87.945}        & 0.223                            \\ \hline
\multicolumn{1}{c|}{P07}  & \multicolumn{1}{c|}{CODA, H}        & \multicolumn{1}{c|}{M}               & 10            & \multicolumn{1}{c|}{87.884}        & \multicolumn{1}{c|}{91.705}        & \multicolumn{1}{c|}{93.641}        & \multicolumn{1}{c|}{94.628}        & \multicolumn{1}{c|}{95.187}        & 0.162                            \\ \hline
\multicolumn{1}{c|}{P08}  & \multicolumn{1}{c|}{Deaf}           & \multicolumn{1}{c|}{M}               & 10            & \multicolumn{1}{c|}{77.636}        & \multicolumn{1}{c|}{83.739}        & \multicolumn{1}{c|}{86.901}        & \multicolumn{1}{c|}{88.662}        & \multicolumn{1}{c|}{89.837}        & 0.211                            \\ \hline
\multicolumn{1}{c|}{P09}  & \multicolumn{1}{c|}{Deaf}           & \multicolumn{1}{c|}{M}               & 7             & \multicolumn{1}{c|}{90.688}        & \multicolumn{1}{c|}{92.993}        & \multicolumn{1}{c|}{94.397}        & \multicolumn{1}{c|}{95.665}        & \multicolumn{1}{c|}{96.345}        & 0.105                            \\ \hline
\end{tabular}
\end{table*}
\label{fig:performance_tops}

We collected a total of 20,604 fingerspelled words for system evaluation from the nine participants. However, due to technical issues, we lost data from three sessions: the 7th session of P1 (116 words), the 7th session of P7 (116 words), and the 5th session of P8 (116 words).

\subsection{Evaluation Metrics}

For evaluating recognition accuracy, we utilized two primary metrics: Letter Error Rate (LER) and word-level accuracy. LER measures the percentage of incorrect letters in the output compared to the ground truth, with a lower value indicating better accuracy. Word-level accuracy was assessed based on top-1 to top-5 predictions, providing a comprehensive view of the model's effectiveness.

\subsubsection{Letter Error Rate (LER)}

LER is a metric used to evaluate the accuracy of a system in recognizing or generating sequences of letters, such as in speech recognition, handwriting recognition, or fingerspelling recognition. It measures the percentage of incorrect letters in the output compared to a reference or ground truth. A lower LER indicates better accuracy in recognizing or generating a sequence of letters. We calculate LER before word correction. For example, when the target word is "hello" and the predicted word from the model is "helo", there is no substitution, one deletion (the second "l" is missing in the predicted word), and no insertion. In this case, the LER is 1/5 = 0.2.

\begin{equation}
LER = \frac{{\text{Substitutions} + \text{Deletions} + \text{Insertions}}}{{\text{Total number of letters in the reference sequence}}}
\end{equation}

\subsubsection{Top N Word-level Accuracy}
We also report word-level accuracy based on top-1 and top-5 predictions, as this provides a more comprehensive evaluation of the model's effectiveness. After collecting the word data, we calculate performance by determining accuracy, defined as the number of correctly predicted words divided by the total number of words. Additionally, we identify the top-3 potential words by selecting those with the smallest Levenshtein distances. For top-3 accuracy, if the correct word is among the three with the smallest Levenshtein distances, the prediction is considered correct; otherwise, it is incorrect. For example, if the input word is "fax" and the model predicts "aax", the three closest words by Levenshtein distance could be "fox", "tax", and "fax." In this case, the system's prediction would be considered correct.

\subsection{Results}

Our analysis revealed that SpellRing achieved a promising overall accuracy of 89.89\% (SD = 8.59\%) for top-1 predictions and 95.72\% (SD = 5.28\%) for top-5 predictions in recognizing the 1,164 fingerspelled words in our dataset. This performance is comparable to \textcolor{black}{FingerSpeller \cite{martin2023fingerspeller} using two rings} (87\% accuracy), demonstrating the effectiveness of our single-ring approach.

\subsubsection{Top-N Word Recognition}

SpellRing's recognition accuracy improved with increasing N in top-N predictions. For top-1 predictions, the system achieved 89.89\% (SD = 8.59\%) accuracy. This improved to 92.85\% (SD = 7.00\%) for top-2, 94.37\% (SD = 6.13\%) for top-3, 95.20\% (SD = 5.56\%) for top-4, and reached 95.72\% (SD = 5.28\%) for top-5 predictions.
Table 3 illustrates the top-N word accuracy for each participant. The significant improvement from top-1 to top-5 accuracy (a 5.83\% increase) suggests potential benefits for ASL translation applications. By considering multiple top predictions, the system could leverage additional contextual information to produce more coherent and contextually appropriate sentences. This approach could help maintain the flow and meaning of the text by selecting from the best few options at each step. Furthermore, a user interface displaying the top three candidates immediately after prediction would allow signers to choose the correct option, potentially improving overall system accuracy.

\begin{figure}[b]
  \includegraphics[width=0.8\linewidth]{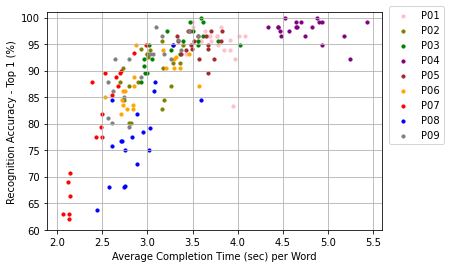}
  \caption{Offline Word-level Performance over Completion Time in User study 1. Dots indicate each session. Note that the model tended to have lower performance for faster signers and better performance for slower signers.}
  \Description{Offline Word-level Performance over Completion Time in User study 1. Dots indicate each sessions. Note that faster signers tended to have lower performance, while slower signers tended to perform better.}
  \label{fig:performance_time}
\end{figure}

\subsubsection{ASL Learners vs. Native/ Fluent Signers}

We observed significant variations in performance across participants, with top-1 accuracy ranging from 77.63\% to 97.85\%. This variability can be attributed to differences in participants' fingerspelling proficiency, affecting articulatory habits and speed. Notably, the model showed better performance for ASL learners (M = 94.38\%, SD = 4.28\%) than for native/ fluent signers (M = 84.06\%, SD = 9.26\%). We attribute this difference primarily to variations in fingerspelling speed.  ASL learners took longer to complete each session (M = 428.88 sec, SD = 74.24) compared to fluent signers (M = 329.624 sec, SD = 42.54). ASL learners tended to fingerspell more slowly, clearly distinguishing each letter, while native/ fluent signers fingerspelled more quickly, resulting in greater handshape variation for certain letters and blurred transitions between letters. This speed difference posed challenges for the model in accurately recognizing individual letters in rapid sequences. We further analyze the impact of fingerspelling speed on performance in the following section.

\subsubsection{Impact of Fingerspelling Speed on Performance}

Fingerspelling speed varied among participants, leading to differences in task completion times. These speed variations, along with factors such as participants' prior experience signing specific words and breaks taken between sessions, appeared to influence the model's performance (Figure \ref{fig:performance_time}). Specifically, the model had lower performance for faster signers and performed better for slower signers.

We found that faster signing speeds often led to greater handshape variation, potentially impacting our model's performance. For instance, when signed quickly, some letters (e.g., ‘C’, ‘O’, ‘E’, ‘I’) were often not fully articulated. The contrast in fingerspelling between the highest and lowest performing participants illustrates this effect. P05, an ASL learner with 3 years' experience, achieved the highest performance with 97.85\% (SD = 2.07\%) accuracy. They articulated each letter very clearly and exhibited minimal handshape variation, resulting in consistent data. In contrast, P08, a fluent signer, had the lowest performance at 77.64\% (SD = 7.8\%). Their fingerspelling speed varied across sessions and consisted of many allophonic handshape variations.

These observations suggest that as fingerspelling speed increases, individual letters are articulated less fully and/or are coarticulated with neighboring letters, posing a challenge for accurate recognition. This challenge resembles those in early-stage speech recognition systems, which experienced performance drops when speakers spoke too rapidly or with strong accents. We discuss this in more detail in Section \ref{impact_speed}.

\section{Phrase-level Recognition}

\subsection{Purpose and Overview}
Building on the insights from our word-level study, our second investigation aimed to evaluate SpellRing's performance for real-time phrase-level recognition. We sought to understand how the system performs in more natural contexts, how users adapt their signing behavior to real-time feedback, and the effectiveness of language models in improving recognition accuracy. It is worth noting that most prior work \cite{paudyal2017dyfav, mummadi2017real} using wearables does not evaluate fingerspelling recognition continuously in real-time. This real-time performance study was a crucial step in assessing SpellRing's potential for practical, everyday use \textcolor{black}{in comprehensive ASL recognition systems.}

\subsection{Participants}
We recruited 11 participants (4 male, 7 female, mean age = 32.0, SD = 3.88) to evaluate our system, consisting of 8 Deaf individuals, 1 ASL interpreter, and 2 hearing ASL learners. The 8 Deaf participants use ASL as their primary language are fluent ASL signers and fingerspellers. The 2 hearing participants had been learning ASL for 1 and 2 years, respectively, contributing to differences in their fingerspelling proficiency. All participants fingerspelled using their right hand as their dominant hand. Detailed information about participants’ proficiency and background is provided in Table \ref{fig:table2}.

\subsection{Iteration on Hardware Prototype Design}
\begin{figure}[t]
  \includegraphics[width=\linewidth]{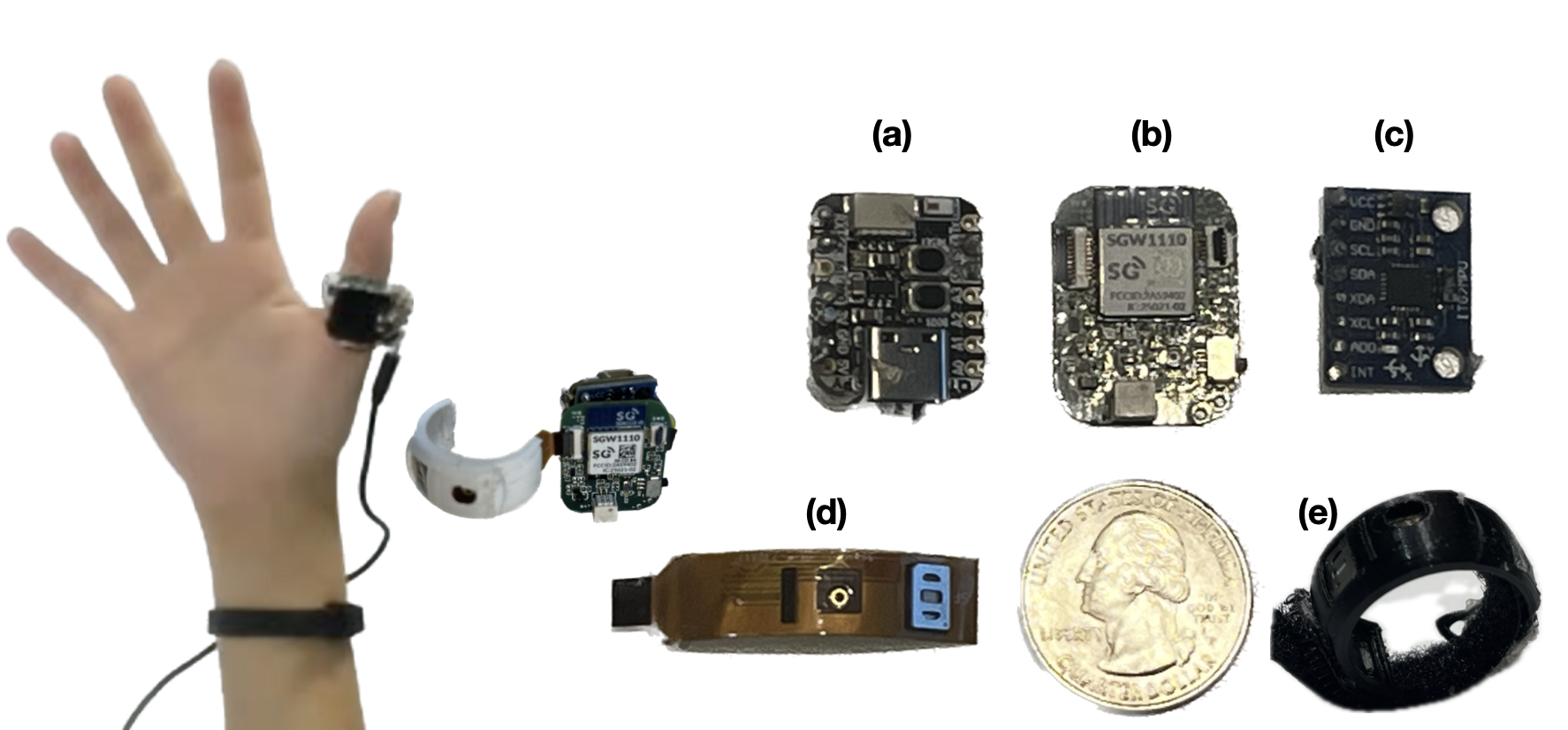}
  \caption{Prototype for real-time phrase-level evaluation: (a) ESP32 S3 microcontroller, (b) nRF MCU, (c) IMU sensor, (d) FPCB, and (e) a 3D-printed ring-shaped design}
  \Description{Prototype for real-time phrase-level evaluation: (a) ESP32 S3 microcontroller, (b) nRF MCU, (c) IMU sensor, (d) FPCB, and (e) a 3D-printed ring-shaped design}
  \label{fig:prototype2}
\end{figure}

To evaluate our system in a more natural fingerspelling environment, we redesigned the ring prototype to a smaller form factor, enabling real-time evaluation. Our design optimized the device for comfort and ease of use while ensuring it supported continuous and real-time tracking for natural fingerspelling. \textcolor{black}{As shown in Figure \ref{fig:prototype2}, audio data from the FPCB microphone \textcolor{black}{(d)} connected to our custom nRF MCU \textcolor{black}{(b)} and gyro data from the IMU \textcolor{black}{(c)} were synchronized and sent to an off-the-shelf ESP32 S3 microcontroller \footnote{Adafruit QT Py ESP32-S3 WiFi Dev Board with STEMMA QT - 8 MB Flash} \textcolor{black}{(a)}.} This data was then transmitted via the wire to a backend system for processing through our machine learning pipeline, running on a MacBook Pro. The raw predictions were processed through autocorrection and language model pipelines to generate the final output.

\subsection{Language Model}
With our dataset, we used an N-gram language model to correct fingerspelled words within phrases. Based on the LM model described in \cite{zhu2018typing}, we generated a list of top N words (N = 20) along with their similarity values after autocorrecting a raw predicted word. For each new predicted word in the top N, we applied bigram and trigram probabilities and selected the word set with the highest probability for the final predicted phrase. We then calcuated WER between the ground truth phrases and the final predicted phrases for evaluation.

\subsection{Dataset and Procedure}

For our phrase-level prediction evaluation, we again utilized the MacKenzie-Soukoreff Phrase Set \cite{mackenzie2003phrase}, as in the first user study. The phrases ranged from 16 to 40 characters in length, consisting of 4 to 8 words each. Our study procedure consisted of two main phases: initial data collection and real-time evaluation.

\subsubsection{Phase 1: Training Data Collection}
We first collected training data from all 11 participants, following a procedure similar to Study 1. Each participant provided two rounds of training data for 1,164 words (2*1164). Our training process involves a two-step
approach: first training with data from all participants except one, then retraining with the specific participant’s data
 for real-time phrase evaluation. Each participant provided the two rounds of training data over two separate days. 
\begin{figure}[t]
  \includegraphics[width=0.8\linewidth]{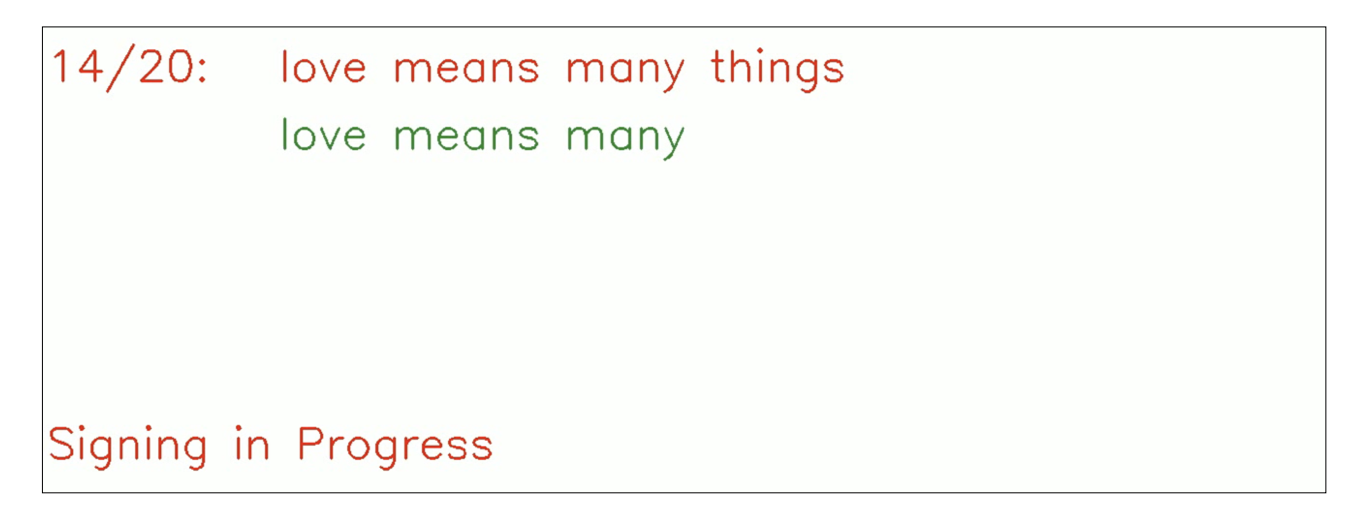}
  \caption{User Interface for Real-time Evaluation: The ground truth phrases are displayed in red, and the predicted phrases appear based on each fingerspelled word. Participants receive status updates below, such as `start signing' and `processing.'  }
  \Description{ser Interface for Real-time Evaluation: The ground truth phrases are displayed in red, and the predicted phrases appear based on each fingerspelled word. Participants receive status updates below, such as 'start signing' and 'processing.'  }
  \label{fig:interface_study2}
\end{figure}
\subsubsection{Phase 2: Real-time Evaluation}
\begin{table*}[t]
\caption{Offline evaluation of 1,164 word-level recognition with WPM, Top N and LER, and real-time evaluation of 100 phrase-level recognition with WER and WPM, G: Gender, H: Hearing, DHH: Deaf or Hard of Hearing. Year: Indicates ASL experience (years learning ASL or using it as a primary language)}
\Description{Offline evaluation of 1,164 word-level recognition with WPM, Top N and LER, and real-time evaluation of 100 phrase-level recognition with WER and WPM,G: Gender, H: Hearing, DHH: Deaf or Hard of Hearing. Year: Indicates ASL experience (years learning ASL or using it as a primary language)}
\begin{tabular}{cccc|ccccccc|cc}
\hline
\textbf{}                      & \textbf{}                           & \textbf{}                            & \textbf{}     & \multicolumn{7}{c|}{Offline}                                                                                                                                                                                                                      & \multicolumn{2}{c}{Real-Time}                       \\
\textbf{}                      & \textbf{}                           & \textbf{}                            & \textbf{}     & \multicolumn{7}{c|}{\textbf{Word-level}}                                                                                                                                                                                                          & \multicolumn{2}{c}{\textbf{Phrase-level}}           \\ \hline
\multicolumn{1}{c|}{\textbf{}} & \multicolumn{1}{c|}{\textbf{Types}} & \multicolumn{1}{c|}{\textbf{G}} & \textbf{Year} & \multicolumn{1}{c|}{\textbf{Top 1}} & \multicolumn{1}{c|}{\textbf{Top 2}} & \multicolumn{1}{c|}{\textbf{Top 3}} & \multicolumn{1}{c|}{\textbf{Top 4}} & \multicolumn{1}{c|}{\textbf{Top 5}} & \multicolumn{1}{c|}{\textbf{LER}}   & \textbf{WPM}  & \multicolumn{1}{c|}{\textbf{WER}}   & \textbf{WPM}  \\ \hline
\multicolumn{1}{c|}{\textcolor{black}{Avg.}}       & \multicolumn{1}{c|}{}               & \multicolumn{1}{c|}{}                & \textit{11}   & \multicolumn{1}{c|}{\textit{82.60}} & \multicolumn{1}{c|}{\textit{87.45}} & \multicolumn{1}{c|}{\textit{90.19}} & \multicolumn{1}{c|}{\textit{91.56}} & \multicolumn{1}{c|}{\textit{92.54}} & \multicolumn{1}{c|}{\textit{0.149}} & \textit{39.9} & \multicolumn{1}{c|}{\textit{0.099}} & \textit{20.1} \\ \hline
\multicolumn{1}{c|}{P01}       & \multicolumn{1}{c|}{Deaf}           & \multicolumn{1}{c|}{M}               & 10            & \multicolumn{1}{c|}{84.87}          & \multicolumn{1}{c|}{89.17}          & \multicolumn{1}{c|}{92.43}          & \multicolumn{1}{c|}{93.64}          & \multicolumn{1}{c|}{94.58}          & \multicolumn{1}{c|}{0.127}          & 32.1          & \multicolumn{1}{c|}{0.041}          & 22.3          \\ \hline
\multicolumn{1}{c|}{P02}       & \multicolumn{1}{c|}{DHH}            & \multicolumn{1}{c|}{F}               & 5             & \multicolumn{1}{c|}{87.21}          & \multicolumn{1}{c|}{90.54}          & \multicolumn{1}{c|}{93.03}          & \multicolumn{1}{c|}{94.84}          & \multicolumn{1}{c|}{95.87}          & \multicolumn{1}{c|}{0.122}          & 39.5          & \multicolumn{1}{c|}{0.124}          & 19.8          \\ \hline
\multicolumn{1}{c|}{P03}       & \multicolumn{1}{c|}{Deaf}           & \multicolumn{1}{c|}{F}               & 37            & \multicolumn{1}{c|}{77.49}          & \multicolumn{1}{c|}{83.33}          & \multicolumn{1}{c|}{87.03}          & \multicolumn{1}{c|}{88.66}          & \multicolumn{1}{c|}{90.29}          & \multicolumn{1}{c|}{0.192}          & 47.9          & \multicolumn{1}{c|}{0.112}          & 20.3          \\ \hline
\multicolumn{1}{c|}{P04}       & \multicolumn{1}{c|}{Deaf}           & \multicolumn{1}{c|}{F}               & 10            & \multicolumn{1}{c|}{77.03}          & \multicolumn{1}{c|}{82.79}          & \multicolumn{1}{c|}{85.62}          & \multicolumn{1}{c|}{87.43}          & \multicolumn{1}{c|}{88.38}          & \multicolumn{1}{c|}{0.193}          & 52.5          & \multicolumn{1}{c|}{0.103}          & 19.7          \\ \hline
\multicolumn{1}{c|}{P05}       & \multicolumn{1}{c|}{Deaf}           & \multicolumn{1}{c|}{M}               & 5             & \multicolumn{1}{c|}{94.40}          & \multicolumn{1}{c|}{97.41}          & \multicolumn{1}{c|}{98.13}          & \multicolumn{1}{c|}{98.71}          & \multicolumn{1}{c|}{98.99}          & \multicolumn{1}{c|}{0.062}          & 34.5          & \multicolumn{1}{c|}{0.061}          & 20.2          \\ \hline
\multicolumn{1}{c|}{P06}       & \multicolumn{1}{c|}{Deaf}           & \multicolumn{1}{c|}{F}               & 22            & \multicolumn{1}{c|}{67.18}          & \multicolumn{1}{c|}{76.63}          & \multicolumn{1}{c|}{79.96}          & \multicolumn{1}{c|}{82.00}          & \multicolumn{1}{c|}{83.82}          & \multicolumn{1}{c|}{0.234}          & 56.8          & \multicolumn{1}{c|}{0.134}          & 24.9          \\ \hline
\multicolumn{1}{c|}{P07}       & \multicolumn{1}{c|}{Deaf}           & \multicolumn{1}{c|}{F}               & 5             & \multicolumn{1}{c|}{90.93}          & \multicolumn{1}{c|}{93.76}          & \multicolumn{1}{c|}{95.65}          & \multicolumn{1}{c|}{96.42}          & \multicolumn{1}{c|}{96.85}          & \multicolumn{1}{c|}{0.1}            & 29.4          & \multicolumn{1}{c|}{0.093}          & 21.3          \\ \hline
\multicolumn{1}{c|}{P08}       & \multicolumn{1}{c|}{Deaf}           & \multicolumn{1}{c|}{M}               & 17            & \multicolumn{1}{c|}{65.27}          & \multicolumn{1}{c|}{72.53}          & \multicolumn{1}{c|}{77.21}          & \multicolumn{1}{c|}{80.17}          & \multicolumn{1}{c|}{81.59}          & \multicolumn{1}{c|}{0.26}           & 52.4          & \multicolumn{1}{c|}{0.17}           & 17.9          \\ \hline
\multicolumn{1}{c|}{P09}       & \multicolumn{1}{c|}{Intepreter, H}  & \multicolumn{1}{c|}{F}               & 10            & \multicolumn{1}{c|}{85.94}          & \multicolumn{1}{c|}{89.67}          & \multicolumn{1}{c|}{93.09}          & \multicolumn{1}{c|}{93.37}          & \multicolumn{1}{c|}{94.86}          & \multicolumn{1}{c|}{0.119}          & 30.1          & \multicolumn{1}{c|}{-}              & -             \\ \hline
\multicolumn{1}{c|}{P10}       & \multicolumn{1}{c|}{Learner, H}     & \multicolumn{1}{c|}{M}               & 2             & \multicolumn{1}{c|}{86.41}          & \multicolumn{1}{c|}{91.92}          & \multicolumn{1}{c|}{94.24}          & \multicolumn{1}{c|}{95.53}          & \multicolumn{1}{c|}{95.95}          & \multicolumn{1}{c|}{0.129}          & 40.2          & \multicolumn{1}{c|}{0.052}          & 20.3          \\ \hline
\multicolumn{1}{c|}{P11}       & \multicolumn{1}{c|}{Learner, H}     & \multicolumn{1}{c|}{F}               & 1             & \multicolumn{1}{c|}{91.92}          & \multicolumn{1}{c|}{94.15}          & \multicolumn{1}{c|}{95.70}          & \multicolumn{1}{c|}{96.38}          & \multicolumn{1}{c|}{96.73}          & \multicolumn{1}{c|}{0.098}          & 23.1          & \multicolumn{1}{c|}{0.101}          & 14.1          \\ \hline
\end{tabular}
\end{table*}
\label{fig:table2}
The real-time evaluation was conducted on a third day.  In the real-time evaluation, we randomly selected 200 phrases generated from the 1164 unique words for our study. We used 100 phrases for practice and the remaining 100 phrases for testing. 

We began with practice sessions, where participants were given 100 phrases to familiarize themselves with our interface (Fig. \ref{fig:interface_study2}) and the real-time feedback mechanism. This preparatory step ensured that participants were comfortable with the system before the actual evaluation. For the evaluation, we used the remaining 100 phrases. Participants fingerspelled these phrases over the course of 5 sessions in natural environments such as their homes or quiet rooms. During each evaluation session, participants fingerspelled according to the prompts illustrated in Figure \ref{fig:interface_study2}. 

Participants were first shown the phrase, and instructed to begin fingerspelling each word after pressing the space key. As they fingerspelled, the real-time prediction model provided immediate feedback by displaying the predicted output in green on the interface.  Participants were instructed to proceed to the next word even if they saw a mispredicted word on the screen to ensure that they continued to fingerspell each word as displayed. The language model sometimes corrected a mispredicted word as signers completed more words in each phrase.

After completing each phrase, participants pressed the space key with their non-dominant hand to display the next phrase, allowing them to see and prepare for it. Once ready, they pressed the key to start the phrase and pressed it again upon completion. This action served a dual purpose: it advanced the interface to the next phrase and also marked the start and end times for fingerspelling. This timing information allowed us to calculate fingerspelling speed for each phrase by minimizing perception time, allowing for more precise estimation of participants' fingerspelling speed and any adjustments they made in response to real-time feedback. The duration of each session varied based on the participant's fingerspelling proficiency and typically ranged from 10 to 12 minutes. This variation in session length allowed us to accommodate different signing speeds and ensure that all participants could complete the phrases comfortably.

In total, we collected data on 993 phrases from 10 participants. Due to technical issues, we lost data for seven phrases, and one participant was unable to complete the entire study. Despite these minor setbacks, the collected data provided a robust basis for evaluating our system's performance in real-time, continuous fingerspelling recognition. This two-phase approach allowed us to first train our system on participant-specific data and then evaluate its performance in a realistic, real-time scenario.

\subsection{Evaluation Metrics}
To evaluate our system, we use Word Error Rate (WER) to report performance. The WER metric ranges from 0 to 1, where 0 indicates that the compared texts are identical, and 1 indicates that they are completely different with no similarity. For example, a WER of 0.10 means there is a 10\% error rate in the compared sentences. WER is based on Levenshtein distance, but it operates at the word level instead of the phoneme (or in this case, letter) level.

\begin{equation}
WER = \frac{{\text{Substitutions} + \text{Deletions} + \text{Insertions}}}{{\text{Total number of words in the reference phrase}}}
\end{equation}

\subsection{Results}
Overall, \theDevice{} recognized 100 phrases with a WER of 0.099\% (0.039\%). While word-level performance achieved an average LER of 0.149\%, phrase-level WER improved with use of a language model.

\subsubsection{Recognition Performance}
Our results show that fingerspelled words are better recognized within the context of a phrase using a language model (See Table 4). The model showed lower performance for faster signers, such as P06 and P08, with a top-1 accuracy of 67.27\% and 65\%, and WERs of 0.134 and 0.17; this translates to approximately 15\% error on the phrases. Compared to word-level recognition performance, this offers an improvement in recognizing fingerspelled words \textcolor{black}{by applying corrections using a language model at phrase-level recognition.}

\subsubsection{Signing Speed in Phrase-Level Prediction}

We observed that participants adjusted their signing speed and habits according to the predictions displayed on the user interface, leading to a decrease in words per minute (WPM) for these participants, with WPM averages ranging from 39.87 (data collected in training phase) to 20.09 (data collected in real-time phase) (See Table 3). This decrease accounts for latency, including model processing time, participants' reaction times, and participants' fingerspelling more slowly in response to prediction accuracy.

\subsubsection{Qualitative Analysis}
Participants were asked open-ended survey questions regarding their overall experience with \theDevice{} in terms of performance, form factor, and usability. For performance, 8 out of 10 participants reported that the system performed well with the language model, even when phrases were entirely misclassified. They noted that short fingerspelled words (e.g., "a," "I," "am", "be") were not always recognized accurately but could be corrected by the language model when more context was available. However, participants noted that the system did not always work well at first. In these cases, we observed changes in participant behavior based on predicted results; they tended to fingerspell more slowly and distinctly immediately after they encountered recognition errors. \textcolor{black}{Although our offline evaluation demonstrated that the system works reliably without requiring participants to alter their signing behavior, real-time evaluation revealed that participants adjusted their fingerspelling habits dynamically to accommodate the system. Specifically, they slowed their fingerspelling immediately after observing misrecognized words but returned to their natural habits when the system performed accurately.}

After some practice sessions, some participants adjusted their fingerspelling speed and habits to accommodate the system. P01, P02, P05, and P07 stated that they focused on spelling clearly and distinctly without skipping letters. This directly contrasted their natural fingerspelling behavior, which often involved partially articulated letters and quick, seamless transitions between letters.

\section{Discussion and Limitations}

\begin{figure*}[t]
  \includegraphics[width=\linewidth]{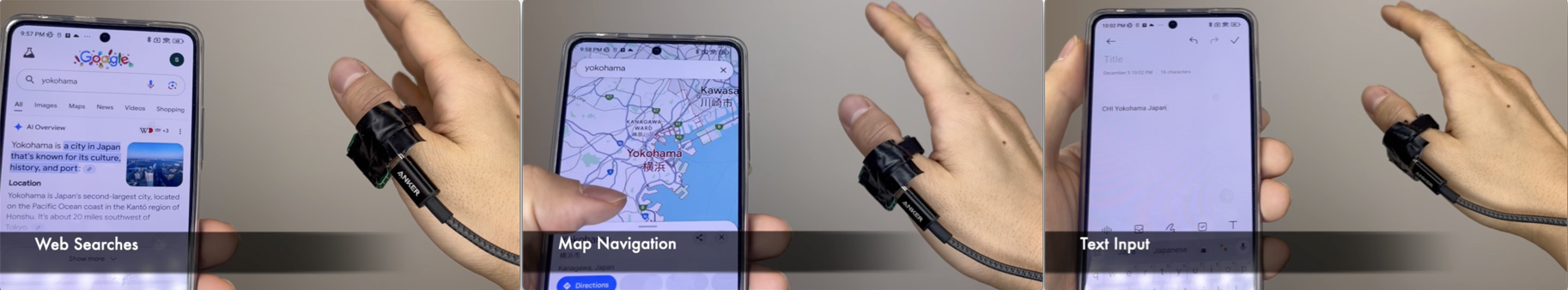}
  \caption{Potential Use cases of SpellRing: Web search, map navigation, and text input on a mobile phone}
  \Description{Potential Use case of SpellRing: Web search, map navigation, and text input on a mobile phone}
  \label{fig:use_case}
\end{figure*}

\textcolor{black}{In this section, we discuss the limitations of SpellRing, followed by the challenges and design implications of developing a ring-based ASL fingerspelling recognition system for potential future use by DHH individuals.}

\subsection{Continuous Finger Tracking vs. Continuous Fingerspelling Recognition}
\textcolor{black}{
Although accurately tracking finger movement is fundamental for fingerspelling recognition, it does not guarantee successful recognition of continuous fingerspelling. Continuous fingerspelling recognition requires an additional layer of linguistic and sequential processing, separating indistinct handshapes and transitions between letters of fingerspelled words. First, finger movement tracking typically focuses on broad, fluid motions without the need to recognize distinct handshapes or subtle transitions between them. Continuous fingerspelling, however, requires high precision to differentiate between similar handshapes, such as those for `M' and `N' in ASL. This requires detailed recognition of finger positions and transitions, meaning that recognition performance can vary depending on the accuracy of the tracking \cite{taylor2018real}. 
Second, fingerspelling involves interpreting static handshapes as letters in sequence to form words. Misrecognizing even a single letter can change the entire meaning of the misrecognized word or phrase, highlighting the critical need for high accuracy and use of a language model to correct errors. In contrast, finger movement tracking does not require the same level of interpretation and can often tolerate minor errors in finger positioning. Finally, continuous fingerspelling recognition requires advanced algorithms, such as Connectionist Temporal Classification (CTC), to parse specific handshape sequences since it allows the system to recognize fingerspelling continuously without needing to label each fingerspelled letter. In contrast, finger movement tracking often relies on motion patterns or positions and lacks the nuanced training necessary to distinguish different handshapes of ASL letters and the transitions between them.
}

\subsection{Extending SpellRing for Large-Scale Fingerspelling }

In this study, we demonstrated how a single ring can recognize 1,164 fingerspelled words across 500 phrases using the MacKenzie dataset, representing a significant advancement over prior wearable-based ASL fingerspelling recognition systems. \textcolor{black}{ While the MacKenzie dataset provides an effective prototype for input systems, enabling us to estimate performance and potential, it also underscores the limitations of our system. We acknowledge that 1,164 words represent a small fraction of the vocabulary compared to commonly used fingerspelled words \cite{shi2018american}, including proper nouns, names, and specific terms essential for ASL conversational vocabulary.}
Since a more extensive vocabulary would encompass a broader range of fingerspelling behaviors and transitions within words, our system could benefit from a significantly larger dataset, which we believe extends beyond the scope of a single research paper. 

Our experiments, however, revealed that pre-training a model on data from other participants and fine-tuning it with the target participant's data significantly improved performance. We evaluated \theDevice{} using a two-step training model, incorporating data from all participants to explore how leveraging cross-participant information could enhance system performance. For comparison, we tested the model using each participant’s individual data with leave-one-session-out cross-validation. The results from both Study 1 and Study 2 showed a top-1 accuracy of 77.21\% (SD=8.94) when trained on individual data, while the pre-trained model with data from other participants improved performance by 5.24\%, achieving 82.45\%. This improvement suggests that cross-participant data enhances the model's ability to generalize across users, leading to better accuracy. This finding suggests a promising approach for collecting a large-scale dataset for pre-trained models, reducing the amount of training data needed per signer. Thus, our system validation could be extended from text input to conversational fingerspelled words, forming part of a comprehensive ASL recognition system.

\subsection{Potential Uses Cases of \theDevice{}}
\textcolor{black}{
ASL fingerspelling, as noted in related work \cite{martin2023fingerspeller, hassan2023tap}, can serve as a fast and accessible text entry tool, outperforming virtual keyboards. Specifically, ASL fingerspelling can be significantly faster than typing on a smartphone’s virtual keyboard \footnote{https://www.kaggle.com/competitions/asl-fingerspelling} \cite{hassan2023tap, martin2023fingerspeller}. In this paper, we highlighted SpellRing’s potential as a text input tool, which could be suitable for devices such as mobile phones, home assistants, and VR/AR glasses, addressing challenges like the restricted camera field of view on these devices. We demonstrated potential applications, including web searches, map navigation, and text entry (as illustrated in Figure \ref{fig:use_case}).}

\textcolor{black}{
We emphasized that while SpellRing is not a comprehensive ASL recognition system and focuses solely on recognizing fingerspelled sequences, its ability to distinguish between different handshapes, palm orientations, and movements in continuous fingerspelling represents a foundational step toward developing a comprehensive ASL recognition system. Building on prior work, such as SignRing \cite{li2023signring}—which uses a single embedded IMU to classify signs—our system shows potential to enhance ASL sentence-level recognition by combining sign classification and fingerspelling recognition, especially in contexts where signs and fingerspelled words are used together (e.g., "MY NAME D-A-V-I-D"). This ability suggests a promising direction for future research and development.
}

\subsection{Auto-Correction and Language Model}\label{auto_correction}

\textcolor{black}{We evaluated our system using the MacKenzie-Soukoreff phrase set \cite{mackenzie2003phrase} to align with previous work \cite{martin2023fingerspeller}, allowing us to estimate performance and potential. Since our auto-correction and n-gram model were developed with our dataset, we conducted additional analysis of auto-correction to assess the generalization of our system by incorporating different reference sets and utilizing large language models. When adding the ChicagoFSWild sets \cite{shi2019fingerspelling, shi2018american} to the reference for auto-correction, our model's top-1 word-level accuracy decreased by 7.1\%, with a smaller drop of 2.5\% in top-5 accuracy, while the word error rate (WER) increased slightly from 0.099 to 0.108. Using a general large language model (LLM), i.e., Llama 3.3 \footnote{https://www.llama.com/}, for phrase-level correction, WER improved to 0.142, demonstrating the potential of language models to mitigate the decline in word-level accuracy. Since the word dictionary used for reference affects performance, we believe that SpellRing could be improved with customized word lists, such as user-defined lists based on their own common language usage, including names and colloquialisms. These lists can be created from their mobile messages or by allowing users to add words themselves.}

\subsection{Design Optimization for Form Factors}

Our system currently utilizes a set of miniature sensors, including a microphone, speaker, and IMU sensor, but as a research prototype, these components are not yet fully optimized in terms of hardware design. For instance, the IMU sensor is housed on a separate PCB board, rather than being integrated with the microcontroller and acoustic sensors on a single, compact PCB. Despite these design limitations, the prototype performed well in initial user studies. However, further hardware optimization is essential to improve its long-term wearability and user comfort. A potential next step would involve fully integrating all sensors and the microcontroller into a single flexible PCB, which could be powered by a curved battery, similar to the compact designs used in commercially available smart rings. This integration would not only streamline the design but also reduce the system's size and improve its overall form factor, making it less obtrusive for everyday use. With these advancements, we envision the prototype evolving into a fully functional, minimally obtrusive ring system that closely resembles off-the-shelf smart rings, offering enhanced practicality and user-friendliness for long-term use.

\subsection{Impact of Speed, Variation, and Clarity on Performance} \label{impact_speed}
From our results, we observed a strong correlation between signing speed and performance, as shown in Figure \ref{fig:performance_time}. This relationship stems from differences in individual signing habits, irrespective of whether or not the signer is Deaf. When signers fingerspell quickly, they do not always articulate each letter clearly, which affects the clarity of certain letters within words. For instance, the letter `E' exhibits considerable variation, especially when positioned in the middle of a word that is fingerspelled rapidly, which poses challenges for the model to accurately learn this letter. However, this variability in performance is largely dependent on individual signers. For instance, the model performed at different levels for fluent signers P7 and P8 from Study 1, stemming from key differences in how clearly and distinctly the participant fingerspelled each letter. Previous work \cite{keane2016fingerspelling} also highlights variation in fingerspelling beyond handshape, including speed and transitions. From Study 2, we gained insight into how signers adjusted their fingerspelling habits based on real-time feedback from the system. For example, they adjusted their fingerspelling speed and clarity to accommodate the system's performance. Signers also reported that for real-time systems, they made an effort not to miss any letters to achieve better performance, whereas in a natural fingerspelling context, signers may have omitted certain letters.


\subsection{User-Independence Performance} \label{pretrained model}
Given the success of Ring-a-Pose \cite{yu2024ring} in achieving strong user-independent performance, \textcolor{black}{we anticipated that our system would also perform well in a user-independent setting. To evaluate this, we conducted a leave-one-participant-out assessment. \textcolor{black}{However, the overall performance achieved was 48.42\% (SD = 12.38), ranging from 32.1\% to 72.13\%, likely due to variations in signers' habits such as fast fingerspelling speed, blurred transitions, and individual fingerspelling styles; the system's performance reflects the real-world complexity of continuous fingerspelling recognition and accounting for variations in users' natural behaviors.}}

{Nevertheless, we recognize the potential of our system, as five participants with slightly slower and clearer signing habits achieved over 63\% accuracy using a user-independent model.} We believe that further expanding the dataset with more diverse participants could yield even greater gains. A larger dataset would likely enhance the model's ability to generalize across different users and recognize more complex actions, gestures, movements, or handshapes. Future work will focus on expanding the dataset and refining the model to optimize its effectiveness in both user-dependent and user-independent scenarios, providing a deeper understanding of how best to enhance \theDevice{}'s performance. Additionally, we expect that future work on transfer learning will help lower the barrier and enhance the usability of a user-independent recognition model.

\subsection{Combining IMU and Acoustic Data for Improving Performance}
We experimented with a sensor fusion approach for ASL fingerspelling recognition by combining acoustic sensing and IMU (Inertial Measurement Unit) data. To evaluate the effectiveness of this multimodal approach, we tested our model using each participant's data from Study 1 and 2 with leave-one-session-out cross-validation under three different conditions: acoustic-only, IMU-only, and sensor fusion. 
Overall, the results showed that the acoustic-only model achieved an accuracy of 72.42\% (SD=12.11\%), while the IMU-only model performed slightly better with 74.29\% (SD=9.22\%). The sensor fusion model, which integrated both data sources, achieved the highest accuracy at 78.11\% (SD=7.88\%), indicating that the multimodal approach outperformed using either modality alone. \textcolor{black}{We note that across all sessions per participant in Study 1 and 2, the sensor fusion model outperformed the other two models. Additionally, the large standard deviation in performance observed across participants was primarily due to variations in their signing speeds rather than the ablation studies.} The fusion model was particularly effective at resolving commonly misclassified letter pairs, such as `P' and `K', `G' and `Q', and `M' and `N', which were frequently confused by the single-modality models. These results suggest that sensor fusion can improve the accuracy and robustness of ASL fingerspelling recognition by leveraging the complementary strengths of both acoustic and IMU data. \textcolor{black}{Nevertheless, while adding IMU sensors proved helpful, we acknowledge that incorporating IMU data did not significantly improve performance. We assume that the acoustic sensor captures some movement information; native and fluent signers fingerspell rapidly, and distinct signals are generated during transitions between letters. We believe this insight will contribute to future design implications of ring-based ASL recognition systems.}

\subsection{Environmental Noise}
\textcolor{black}{Our multimodal approach, utilizing both acoustic and IMU data, enhances the system's robustness against environmental noise, particularly since the IMU sensor is unaffected by such noise. Our ablation study demonstrated that the system performs reliably using a single IMU only but achieves better performance when acoustic signals are included. Conducted in semi-controlled environments, ranging from quiet rooms to participants' homes with background noise from roommates, our study showed that the system delivers reliable performance in everyday noise conditions. Nevertheless, environmental acoustic noise, such as conversations or keyboard typing, can degrade signal quality and affect performance. To mitigate this, we applied band-pass filtering during signal processing to address lower-frequency environmental noise. Drawing on similar work using active acoustic sensing \cite{yu2024ring, lee2024echowrist, zhang2023echospeech, li2022eario}, incorporating environmental noise data into the model could further enhance the system's resilience against acoustic interference.}

\subsection{Future Work}
\label{future}

The primary goal of this paper is to demonstrate the potential of a single, cost-effective ring \textcolor{black}{(with a prototype cost of approximately US\$30 \cite{yu2024ring}, which could decrease with mass production)} for ASL fingerspelling recognition. By sharing our findings, we aim to encourage more researchers in the field to collaborate in further developing this system.  Our ultimate goal is to create an ASL translation system that Deaf and hard of hearing individuals could use in their daily lives to aid in communication between DHH and hearing individuals. 

\textcolor{black}{As we move forward, future work should focus on integrating our system into AR smart glasses, where the entire system can recognize and distinguish subtle differences and individual variations in all ASL phonological parameters (i.e., handshape, palm orientation, movement, location, and non-manual markers) \cite{valli1992linguistics} to advance progress toward ASL translation, sign-to-speech, and/or speech-to-text/sign-on-glass displays for DHH users. Our system's ability to track handshape, palm orientation, and movement could enhance the system's robustness by overcoming the limitations of camera use, such as restricted field-of-view and poor lighting conditions.} Future work should also incorporate tracking of both hands to advance recognition of full ASL signs and sentences, rather than of just fingerspelled words.

Additionally, future work should focus on developing more robust machine learning models capable of handling a larger vocabulary and more varied signing styles while maintaining or improving accuracy. Expanding the dataset, refining algorithms, and incorporating contextual information—such as preceding words or signs—could allow us to leverage language models for more accurate predictions. For example, in a conversation about closed captioning, fingerspelled words related to captioning could be corrected more accurately using statistical approaches.

Most notably, training systems to recognize actual ASL signs (as opposed to just fingerspelled English words), is a crucial next step in working toward developing a wearable ASL translation/ recognition tool. While SpellRing focuses on fingerspelling recognition, systems like SignRing \cite{li2023signring} have shown the potential of ring devices with IMU sensors to recognize ASL signs. Our system, which can track handshape, palm orientation, and movement, could also be expanded to recognize ASL signs that share all properties except for handshape (e.g., FAMILY, CLASS, and TEAM), palm orientation (e.g., MAYBE and BALANCE), or movement (e.g., SIT and CHAIR). This capability could significantly enhance the accuracy and applicability of the system for DHH users. 

Continual feedback from Deaf and hard of hearing (DHH) signers will be essential in ensuring future systems meet user needs and preferences. \textcolor{black}{One example is the need to evaluate preferences on ring placement; as demonstrated by Ring-a-Pose \cite{yu2024ring}, changing ring placement maintains potential for fingerspelling recognition, though it may sacrifice performance to improve user experience.}
As we expand SpellRing’s capabilities, we must strike a balance between technological advancement and user comfort, always prioritizing the user's experience and input while creating a tool that recognizes natural signing data in real-time.

\section{Conclusion}
In this paper, we introduced SpellRing, a smart ring designed to continuously recognize words fingerspelled in ASL using a combination of acoustic sensing and IMU sensors. Our system tracks handshape, movement, and palm orientation, applying a deep learning model with Connectionist Temporal Classification (CTC) for word-level recognition. Through evaluations with 20 ASL signers, including 12 fluent signers and 8 ASL learners, SpellRing achieved promising results, with a top-1 recognition accuracy of 82.45\% and a real-time phrase-level Word Error Rate (WER) of 0.099\%. These results demonstrate the potential of using minimally obtrusive wearable technology for ASL recognition and translation. We also explored the impact of a two-step training model, sensor fusion, and design optimization for improving performance, showing that integrating data from multiple participants significantly enhances recognition accuracy. Future work will focus on expanding the dataset, refining hardware, and incorporating more complex ASL signs to create a comprehensive, user-friendly system for Deaf and hard of hearing individuals.

\bibliographystyle{ACM-Reference-Format}
\bibliography{main}


\begin{thebibliography}{75}


\ifx \showCODEN    \undefined \def \showCODEN     #1{\unskip}     \fi
\ifx \showDOI      \undefined \def \showDOI       #1{#1}\fi
\ifx \showISBNx    \undefined \def \showISBNx     #1{\unskip}     \fi
\ifx \showISBNxiii \undefined \def \showISBNxiii  #1{\unskip}     \fi
\ifx \showISSN     \undefined \def \showISSN      #1{\unskip}     \fi
\ifx \showLCCN     \undefined \def \showLCCN      #1{\unskip}     \fi
\ifx \shownote     \undefined \def \shownote      #1{#1}          \fi
\ifx \showarticletitle \undefined \def \showarticletitle #1{#1}   \fi
\ifx \showURL      \undefined \def \showURL       {\relax}        \fi
\providecommand\bibfield[2]{#2}
\providecommand\bibinfo[2]{#2}
\providecommand\natexlab[1]{#1}
\providecommand\showeprint[2][]{arXiv:#2}

\bibitem[Abreu et~al\mbox{.}(2016)]%
        {abreu2016evaluating}
\bibfield{author}{\bibinfo{person}{Jo{\~a}o~Gabriel Abreu}, \bibinfo{person}{Jo{\~a}o~Marcelo Teixeira}, \bibinfo{person}{Lucas~Silva Figueiredo}, {and} \bibinfo{person}{Veronica Teichrieb}.} \bibinfo{year}{2016}\natexlab{}.
\newblock \showarticletitle{Evaluating sign language recognition using the myo armband}. In \bibinfo{booktitle}{\emph{2016 XVIII symposium on virtual and augmented reality (SVR)}}. IEEE, \bibinfo{pages}{64--70}.
\newblock


\bibitem[Ahmed et~al\mbox{.}(2018)]%
        {ahmed2018review}
\bibfield{author}{\bibinfo{person}{Mohamed~Aktham Ahmed}, \bibinfo{person}{Bilal~Bahaa Zaidan}, \bibinfo{person}{Aws~Alaa Zaidan}, \bibinfo{person}{Mahmood~Maher Salih}, {and} \bibinfo{person}{Muhammad Modi~Bin Lakulu}.} \bibinfo{year}{2018}\natexlab{}.
\newblock \showarticletitle{A review on systems-based sensory gloves for sign language recognition state of the art between 2007 and 2017}.
\newblock \bibinfo{journal}{\emph{Sensors}} \bibinfo{volume}{18}, \bibinfo{number}{7} (\bibinfo{year}{2018}), \bibinfo{pages}{2208}.
\newblock


\bibitem[Aloysius and Geetha(2020)]%
        {aloysius2020understanding}
\bibfield{author}{\bibinfo{person}{Neena Aloysius} {and} \bibinfo{person}{M Geetha}.} \bibinfo{year}{2020}\natexlab{}.
\newblock \showarticletitle{Understanding vision-based continuous sign language recognition}.
\newblock \bibinfo{journal}{\emph{Multimedia Tools and Applications}} \bibinfo{volume}{79}, \bibinfo{number}{31} (\bibinfo{year}{2020}), \bibinfo{pages}{22177--22209}.
\newblock


\bibitem[Ben Haj~Amor et~al\mbox{.}(2023)]%
        {ben2023sign}
\bibfield{author}{\bibinfo{person}{Amina Ben Haj~Amor}, \bibinfo{person}{Oussama El~Ghoul}, {and} \bibinfo{person}{Mohamed Jemni}.} \bibinfo{year}{2023}\natexlab{}.
\newblock \showarticletitle{Sign Language Recognition Using the Electromyographic Signal: A Systematic Literature Review}.
\newblock \bibinfo{journal}{\emph{Sensors}} \bibinfo{volume}{23}, \bibinfo{number}{19} (\bibinfo{year}{2023}), \bibinfo{pages}{8343}.
\newblock


\bibitem[Boh{\'a}{\v{c}}ek and Hr{\'u}z(2022)]%
        {bohavcek2022sign}
\bibfield{author}{\bibinfo{person}{Maty{\'a}{\v{s}} Boh{\'a}{\v{c}}ek} {and} \bibinfo{person}{Marek Hr{\'u}z}.} \bibinfo{year}{2022}\natexlab{}.
\newblock \showarticletitle{Sign pose-based transformer for word-level sign language recognition}. In \bibinfo{booktitle}{\emph{Proceedings of the IEEE/CVF winter conference on applications of computer vision}}. \bibinfo{pages}{182--191}.
\newblock


\bibitem[Chen et~al\mbox{.}(2018)]%
        {chen2018finger}
\bibfield{author}{\bibinfo{person}{Feiyu Chen}, \bibinfo{person}{Jia Deng}, \bibinfo{person}{Zhibo Pang}, \bibinfo{person}{Majid Baghaei~Nejad}, \bibinfo{person}{Huayong Yang}, {and} \bibinfo{person}{Geng Yang}.} \bibinfo{year}{2018}\natexlab{}.
\newblock \showarticletitle{Finger angle-based hand gesture recognition for smart infrastructure using wearable wrist-worn camera}.
\newblock \bibinfo{journal}{\emph{Applied Sciences}} \bibinfo{volume}{8}, \bibinfo{number}{3} (\bibinfo{year}{2018}), \bibinfo{pages}{369}.
\newblock


\bibitem[Du et~al\mbox{.}(2022)]%
        {du2022full}
\bibfield{author}{\bibinfo{person}{Yao Du}, \bibinfo{person}{Pan Xie}, \bibinfo{person}{Mingye Wang}, \bibinfo{person}{Xiaohui Hu}, \bibinfo{person}{Zheng Zhao}, {and} \bibinfo{person}{Jiaqi Liu}.} \bibinfo{year}{2022}\natexlab{}.
\newblock \showarticletitle{Full transformer network with masking future for word-level sign language recognition}.
\newblock \bibinfo{journal}{\emph{Neurocomputing}}  \bibinfo{volume}{500} (\bibinfo{year}{2022}), \bibinfo{pages}{115--123}.
\newblock


\bibitem[Fayyazsanavi et~al\mbox{.}(2024)]%
        {fayyazsanavi2024fingerspelling}
\bibfield{author}{\bibinfo{person}{Pooya Fayyazsanavi}, \bibinfo{person}{Negar Nejatishahidin}, {and} \bibinfo{person}{Jana Ko{\v{s}}eck{\'a}}.} \bibinfo{year}{2024}\natexlab{}.
\newblock \showarticletitle{Fingerspelling PoseNet: Enhancing Fingerspelling Translation with Pose-Based Transformer Models}. In \bibinfo{booktitle}{\emph{Proceedings of the IEEE/CVF Winter Conference on Applications of Computer Vision}}. \bibinfo{pages}{1120--1130}.
\newblock


\bibitem[Feris et~al\mbox{.}(2005)]%
        {feris2005recognition}
\bibfield{author}{\bibinfo{person}{Rogerio Feris}, \bibinfo{person}{Matthew Turk}, \bibinfo{person}{Ramesh Raskar}, \bibinfo{person}{Kar-Han Tan}, {and} \bibinfo{person}{Gosuke Ohashi}.} \bibinfo{year}{2005}\natexlab{}.
\newblock \showarticletitle{Recognition of isolated fingerspelling gestures using depth edges}.
\newblock \bibinfo{journal}{\emph{Real-Time Vision for Human-Computer Interaction}} (\bibinfo{year}{2005}), \bibinfo{pages}{43--56}.
\newblock


\bibitem[Fowley and Ventresque(2021)]%
        {fowley2021sign}
\bibfield{author}{\bibinfo{person}{Frank Fowley} {and} \bibinfo{person}{Anthony Ventresque}.} \bibinfo{year}{2021}\natexlab{}.
\newblock \showarticletitle{Sign Language Fingerspelling Recognition using Synthetic Data.}. In \bibinfo{booktitle}{\emph{AICS}}. \bibinfo{pages}{84--95}.
\newblock


\bibitem[Fujikawa et~al\mbox{.}(2019)]%
        {fujikawa2019development}
\bibfield{author}{\bibinfo{person}{Nami Fujikawa}, \bibinfo{person}{Natsuhiko Hirabayashi}, \bibinfo{person}{Yoshinori Fujisawa}, {and} \bibinfo{person}{Ryohei Yoshimura}.} \bibinfo{year}{2019}\natexlab{}.
\newblock \showarticletitle{Development of Learning Support Equipment for Sign Language and Fingerspelling by Mixed Reality}. In \bibinfo{booktitle}{\emph{Proceedings of the 7th ACIS International Conference on Applied Computing and Information Technology}}. ACM.
\newblock


\bibitem[Gangakhedkar and Theil(2024)]%
        {gangakhedkar2024fingarspell}
\bibfield{author}{\bibinfo{person}{Raj Gangakhedkar} {and} \bibinfo{person}{Arthur Theil}.} \bibinfo{year}{2024}\natexlab{}.
\newblock \showarticletitle{fingARspell: A Mobile AR Tool for Learning the Deafblind Manual Alphabet}. In \bibinfo{booktitle}{\emph{Proceedings of the International Conference on Mobile and Ubiquitous Multimedia}}. \bibinfo{pages}{478--480}.
\newblock


\bibitem[Glasser et~al\mbox{.}(2022)]%
        {glasser2022analyzing}
\bibfield{author}{\bibinfo{person}{Abraham Glasser}, \bibinfo{person}{Matthew Watkins}, \bibinfo{person}{Kira Hart}, \bibinfo{person}{Sooyeon Lee}, {and} \bibinfo{person}{Matt Huenerfauth}.} \bibinfo{year}{2022}\natexlab{}.
\newblock \showarticletitle{Analyzing deaf and hard-of-hearing users’ behavior, usage, and interaction with a personal assistant device that understands sign-language input}. In \bibinfo{booktitle}{\emph{Proceedings of the 2022 CHI Conference on Human Factors in Computing Systems}}. \bibinfo{pages}{1--12}.
\newblock


\bibitem[Goh and Holden(2006)]%
        {goh2006dynamic}
\bibfield{author}{\bibinfo{person}{Paul Goh} {and} \bibinfo{person}{Eun-Jung Holden}.} \bibinfo{year}{2006}\natexlab{}.
\newblock \showarticletitle{Dynamic fingerspelling recognition using geometric and motion features}. In \bibinfo{booktitle}{\emph{2006 International Conference on Image Processing}}. IEEE, \bibinfo{pages}{2741--2744}.
\newblock


\bibitem[Graves et~al\mbox{.}(2006)]%
        {graves2006connectionist}
\bibfield{author}{\bibinfo{person}{Alex Graves}, \bibinfo{person}{Santiago Fern{\'a}ndez}, \bibinfo{person}{Faustino Gomez}, {and} \bibinfo{person}{J{\"u}rgen Schmidhuber}.} \bibinfo{year}{2006}\natexlab{}.
\newblock \showarticletitle{Connectionist temporal classification: labelling unsegmented sequence data with recurrent neural networks}. In \bibinfo{booktitle}{\emph{Proceedings of the 23rd international conference on Machine learning}}. \bibinfo{pages}{369--376}.
\newblock


\bibitem[Graves and Graves(2012)]%
        {graves2012connectionist}
\bibfield{author}{\bibinfo{person}{Alex Graves} {and} \bibinfo{person}{Alex Graves}.} \bibinfo{year}{2012}\natexlab{}.
\newblock \showarticletitle{Connectionist temporal classification}.
\newblock \bibinfo{journal}{\emph{Supervised sequence labelling with recurrent neural networks}} (\bibinfo{year}{2012}), \bibinfo{pages}{61--93}.
\newblock


\bibitem[Gu et~al\mbox{.}(2022)]%
        {gu2022american}
\bibfield{author}{\bibinfo{person}{Yutong Gu}, \bibinfo{person}{Weiyi Wei}, \bibinfo{person}{Xinya Li}, \bibinfo{person}{Jianan Yuan}, {and} \bibinfo{person}{Masahiro Todoh}.} \bibinfo{year}{2022}\natexlab{}.
\newblock \showarticletitle{American Sign Language alphabet recognition using inertial motion capture system with deep learning}.
\newblock \bibinfo{journal}{\emph{Inventions}} \bibinfo{volume}{7}, \bibinfo{number}{4} (\bibinfo{year}{2022}), \bibinfo{pages}{112}.
\newblock


\bibitem[Hanson(1982)]%
        {hanson1982use}
\bibfield{author}{\bibinfo{person}{Vicki~L Hanson}.} \bibinfo{year}{1982}\natexlab{}.
\newblock \showarticletitle{Use of orthographic structure by deaf adults: Recognition of fingerspelled words}.
\newblock \bibinfo{journal}{\emph{Applied Psycholinguistics}} \bibinfo{volume}{3}, \bibinfo{number}{4} (\bibinfo{year}{1982}), \bibinfo{pages}{343--356}.
\newblock


\bibitem[Hassan et~al\mbox{.}(2023)]%
        {hassan2023tap}
\bibfield{author}{\bibinfo{person}{Saad Hassan}, \bibinfo{person}{Abraham Glasser}, \bibinfo{person}{Max Shengelia}, \bibinfo{person}{Thad Starner}, \bibinfo{person}{Sean Forbes}, \bibinfo{person}{Nathan Qualls}, {and} \bibinfo{person}{Sam~S Sepah}.} \bibinfo{year}{2023}\natexlab{}.
\newblock \showarticletitle{Tap to Sign: Towards using American Sign Language for Text Entry on Smartphones}.
\newblock \bibinfo{journal}{\emph{Proceedings of the ACM on Human-Computer Interaction}} \bibinfo{volume}{7}, \bibinfo{number}{MHCI} (\bibinfo{year}{2023}), \bibinfo{pages}{1--23}.
\newblock


\bibitem[Hirabayashi et~al\mbox{.}(2019)]%
        {hirabayashi2019development}
\bibfield{author}{\bibinfo{person}{Natsuhiko Hirabayashi}, \bibinfo{person}{Nami Fujikawa}, \bibinfo{person}{Ryohei Yoshimura}, {and} \bibinfo{person}{Yoshinori Fujisawa}.} \bibinfo{year}{2019}\natexlab{}.
\newblock \showarticletitle{Development of learning support equipment for sign language and fingerspelling by mixed reality}. In \bibinfo{booktitle}{\emph{Proceedings of the 7th ACIS International Conference on Applied Computing and Information Technology}}. \bibinfo{pages}{1--6}.
\newblock


\bibitem[Hosain et~al\mbox{.}(2021)]%
        {hosain2021hand}
\bibfield{author}{\bibinfo{person}{Al~Amin Hosain}, \bibinfo{person}{Panneer~Selvam Santhalingam}, \bibinfo{person}{Parth Pathak}, \bibinfo{person}{Huzefa Rangwala}, {and} \bibinfo{person}{Jana Kosecka}.} \bibinfo{year}{2021}\natexlab{}.
\newblock \showarticletitle{Hand pose guided 3d pooling for word-level sign language recognition}. In \bibinfo{booktitle}{\emph{Proceedings of the IEEE/CVF winter conference on applications of computer vision}}. \bibinfo{pages}{3429--3439}.
\newblock


\bibitem[Hu et~al\mbox{.}(2020)]%
        {hu2020fingertrak}
\bibfield{author}{\bibinfo{person}{Fang Hu}, \bibinfo{person}{Peng He}, \bibinfo{person}{Songlin Xu}, \bibinfo{person}{Yin Li}, {and} \bibinfo{person}{Cheng Zhang}.} \bibinfo{year}{2020}\natexlab{}.
\newblock \showarticletitle{FingerTrak: Continuous 3D hand pose tracking by deep learning hand silhouettes captured by miniature thermal cameras on wrist}.
\newblock \bibinfo{journal}{\emph{Proceedings of the ACM on Interactive, Mobile, Wearable and Ubiquitous Technologies}} \bibinfo{volume}{4}, \bibinfo{number}{2} (\bibinfo{year}{2020}), \bibinfo{pages}{1--24}.
\newblock


\bibitem[Jani et~al\mbox{.}(2018)]%
        {jani2018sensor}
\bibfield{author}{\bibinfo{person}{Abhishek~B Jani}, \bibinfo{person}{Nishith~A Kotak}, {and} \bibinfo{person}{Anil~K Roy}.} \bibinfo{year}{2018}\natexlab{}.
\newblock \showarticletitle{Sensor based hand gesture recognition system for English alphabets used in sign language of deaf-mute people}. In \bibinfo{booktitle}{\emph{2018 IEEE SENSORS}}. IEEE, \bibinfo{pages}{1--4}.
\newblock


\bibitem[Kakoty and Sharma(2018)]%
        {kakoty2018recognition}
\bibfield{author}{\bibinfo{person}{Nayan~M Kakoty} {and} \bibinfo{person}{Manalee~Dev Sharma}.} \bibinfo{year}{2018}\natexlab{}.
\newblock \showarticletitle{Recognition of sign language alphabets and numbers based on hand kinematics using a data glove}.
\newblock \bibinfo{journal}{\emph{Procedia Computer Science}}  \bibinfo{volume}{133} (\bibinfo{year}{2018}), \bibinfo{pages}{55--62}.
\newblock


\bibitem[Keane and Brentari(2016)]%
        {keane2016fingerspelling}
\bibfield{author}{\bibinfo{person}{Jonathan Keane} {and} \bibinfo{person}{Diane Brentari}.} \bibinfo{year}{2016}\natexlab{}.
\newblock \showarticletitle{Fingerspelling: Beyond handshape sequences}.
\newblock \bibinfo{journal}{\emph{The Oxford handbook of deaf studies in language}} (\bibinfo{year}{2016}), \bibinfo{pages}{146--160}.
\newblock


\bibitem[Keane et~al\mbox{.}(2015)]%
        {keane2015segmentation}
\bibfield{author}{\bibinfo{person}{Jonathan Keane}, \bibinfo{person}{Diane Brentari}, {and} \bibinfo{person}{Jason Riggle}.} \bibinfo{year}{2015}\natexlab{}.
\newblock \showarticletitle{Segmentation and pinky extension in ASL fingerspelling}.
\newblock \bibinfo{journal}{\emph{The segment in phonetics and phonology}} (\bibinfo{year}{2015}), \bibinfo{pages}{103--128}.
\newblock


\bibitem[Kim et~al\mbox{.}(2017)]%
        {kim2017lexicon}
\bibfield{author}{\bibinfo{person}{Taehwan Kim}, \bibinfo{person}{Jonathan Keane}, \bibinfo{person}{Weiran Wang}, \bibinfo{person}{Hao Tang}, \bibinfo{person}{Jason Riggle}, \bibinfo{person}{Gregory Shakhnarovich}, \bibinfo{person}{Diane Brentari}, {and} \bibinfo{person}{Karen Livescu}.} \bibinfo{year}{2017}\natexlab{}.
\newblock \showarticletitle{Lexicon-free fingerspelling recognition from video: Data, models, and signer adaptation}.
\newblock \bibinfo{journal}{\emph{Computer Speech \& Language}}  \bibinfo{volume}{46} (\bibinfo{year}{2017}), \bibinfo{pages}{209--232}.
\newblock


\bibitem[Kudrinko et~al\mbox{.}(2022)]%
        {kudrinko2022assessing}
\bibfield{author}{\bibinfo{person}{Karly Kudrinko}, \bibinfo{person}{Emile Flavin}, \bibinfo{person}{Michael Shepertycky}, {and} \bibinfo{person}{Qingguo Li}.} \bibinfo{year}{2022}\natexlab{}.
\newblock \showarticletitle{Assessing the need for a wearable sign language recognition device for deaf individuals: Results from a national questionnaire}.
\newblock \bibinfo{journal}{\emph{Assistive Technology}} \bibinfo{volume}{34}, \bibinfo{number}{6} (\bibinfo{year}{2022}), \bibinfo{pages}{684--697}.
\newblock


\bibitem[Kudrinko et~al\mbox{.}(2020)]%
        {kudrinko2020wearable}
\bibfield{author}{\bibinfo{person}{Karly Kudrinko}, \bibinfo{person}{Emile Flavin}, \bibinfo{person}{Xiaodan Zhu}, {and} \bibinfo{person}{Qingguo Li}.} \bibinfo{year}{2020}\natexlab{}.
\newblock \showarticletitle{Wearable sensor-based sign language recognition: A comprehensive review}.
\newblock \bibinfo{journal}{\emph{IEEE Reviews in Biomedical Engineering}}  \bibinfo{volume}{14} (\bibinfo{year}{2020}), \bibinfo{pages}{82--97}.
\newblock


\bibitem[Lee et~al\mbox{.}(2020)]%
        {lee2020sensor}
\bibfield{author}{\bibinfo{person}{Boon~Giin Lee}, \bibinfo{person}{Teak-Wei Chong}, {and} \bibinfo{person}{Wan-Young Chung}.} \bibinfo{year}{2020}\natexlab{}.
\newblock \showarticletitle{Sensor fusion of motion-based sign language interpretation with deep learning}.
\newblock \bibinfo{journal}{\emph{Sensors}} \bibinfo{volume}{20}, \bibinfo{number}{21} (\bibinfo{year}{2020}), \bibinfo{pages}{6256}.
\newblock


\bibitem[Lee et~al\mbox{.}(2024)]%
        {lee2024echowrist}
\bibfield{author}{\bibinfo{person}{Chi-Jung Lee}, \bibinfo{person}{Ruidong Zhang}, \bibinfo{person}{Devansh Agarwal}, \bibinfo{person}{Tianhong~Catherine Yu}, \bibinfo{person}{Vipin Gunda}, \bibinfo{person}{Oliver Lopez}, \bibinfo{person}{James Kim}, \bibinfo{person}{Sicheng Yin}, \bibinfo{person}{Boao Dong}, \bibinfo{person}{Ke Li}, {et~al\mbox{.}}} \bibinfo{year}{2024}\natexlab{}.
\newblock \showarticletitle{Echowrist: Continuous hand pose tracking and hand-object interaction recognition using low-power active acoustic sensing on a wristband}. In \bibinfo{booktitle}{\emph{Proceedings of the CHI Conference on Human Factors in Computing Systems}}. \bibinfo{pages}{1--21}.
\newblock


\bibitem[Levenshtein(1965)]%
        {Levenshtein1965BinaryCC}
\bibfield{author}{\bibinfo{person}{Vladimir~I. Levenshtein}.} \bibinfo{year}{1965}\natexlab{}.
\newblock \showarticletitle{Binary codes capable of correcting deletions, insertions, and reversals}.
\newblock \bibinfo{journal}{\emph{Soviet physics. Doklady}}  \bibinfo{volume}{10} (\bibinfo{year}{1965}), \bibinfo{pages}{707--710}.
\newblock
\urldef\tempurl%
\url{https://api.semanticscholar.org/CorpusID:60827152}
\showURL{%
\tempurl}


\bibitem[Li et~al\mbox{.}(2023)]%
        {li2023signring}
\bibfield{author}{\bibinfo{person}{Jiyang Li}, \bibinfo{person}{Lin Huang}, \bibinfo{person}{Siddharth Shah}, \bibinfo{person}{Sean~J Jones}, \bibinfo{person}{Yincheng Jin}, \bibinfo{person}{Dingran Wang}, \bibinfo{person}{Adam Russell}, \bibinfo{person}{Seokmin Choi}, \bibinfo{person}{Yang Gao}, \bibinfo{person}{Junsong Yuan}, {et~al\mbox{.}}} \bibinfo{year}{2023}\natexlab{}.
\newblock \showarticletitle{SignRing: Continuous American Sign Language Recognition Using IMU Rings and Virtual IMU Data}.
\newblock \bibinfo{journal}{\emph{Proceedings of the ACM on Interactive, Mobile, Wearable and Ubiquitous Technologies}} \bibinfo{volume}{7}, \bibinfo{number}{3} (\bibinfo{year}{2023}), \bibinfo{pages}{1--29}.
\newblock


\bibitem[Li et~al\mbox{.}(2024a)]%
        {li2024sonicid}
\bibfield{author}{\bibinfo{person}{Ke Li}, \bibinfo{person}{Devansh Agarwal}, \bibinfo{person}{Ruidong Zhang}, \bibinfo{person}{Vipin Gunda}, \bibinfo{person}{Tianjun Mo}, \bibinfo{person}{Saif Mahmud}, \bibinfo{person}{Boao Chen}, \bibinfo{person}{Fran{\c{c}}ois Guimbreti{\v{e}}re}, {and} \bibinfo{person}{Cheng Zhang}.} \bibinfo{year}{2024}\natexlab{a}.
\newblock \showarticletitle{SonicID: User Identification on Smart Glasses with Acoustic Sensing}.
\newblock \bibinfo{journal}{\emph{Proceedings of the ACM on Interactive, Mobile, Wearable and Ubiquitous Technologies}} \bibinfo{volume}{8}, \bibinfo{number}{4} (\bibinfo{year}{2024}), \bibinfo{pages}{1--27}.
\newblock


\bibitem[Li et~al\mbox{.}(2024c)]%
        {li2024gazetrak}
\bibfield{author}{\bibinfo{person}{Ke Li}, \bibinfo{person}{Ruidong Zhang}, \bibinfo{person}{Boao Chen}, \bibinfo{person}{Siyuan Chen}, \bibinfo{person}{Sicheng Yin}, \bibinfo{person}{Saif Mahmud}, \bibinfo{person}{Qikang Liang}, \bibinfo{person}{Fran{\c{c}}ois Guimbreti{\`e}re}, {and} \bibinfo{person}{Cheng Zhang}.} \bibinfo{year}{2024}\natexlab{c}.
\newblock \showarticletitle{GazeTrak: Exploring Acoustic-based Eye Tracking on a Glass Frame}. In \bibinfo{booktitle}{\emph{Proceedings of the 30th Annual International Conference on Mobile Computing and Networking}}. \bibinfo{pages}{497--512}.
\newblock


\bibitem[Li et~al\mbox{.}(2024b)]%
        {li2024eyeecho}
\bibfield{author}{\bibinfo{person}{Ke Li}, \bibinfo{person}{Ruidong Zhang}, \bibinfo{person}{Siyuan Chen}, \bibinfo{person}{Boao Chen}, \bibinfo{person}{Mose Sakashita}, \bibinfo{person}{Fran{\c{c}}ois Guimbreti{\`e}re}, {and} \bibinfo{person}{Cheng Zhang}.} \bibinfo{year}{2024}\natexlab{b}.
\newblock \showarticletitle{EyeEcho: Continuous and Low-power Facial Expression Tracking on Glasses}. In \bibinfo{booktitle}{\emph{Proceedings of the CHI Conference on Human Factors in Computing Systems}}. \bibinfo{pages}{1--24}.
\newblock


\bibitem[Li et~al\mbox{.}(2022)]%
        {li2022eario}
\bibfield{author}{\bibinfo{person}{Ke Li}, \bibinfo{person}{Ruidong Zhang}, \bibinfo{person}{Bo Liang}, \bibinfo{person}{Fran{\c{c}}ois Guimbreti{\`e}re}, {and} \bibinfo{person}{Cheng Zhang}.} \bibinfo{year}{2022}\natexlab{}.
\newblock \showarticletitle{Eario: A low-power acoustic sensing earable for continuously tracking detailed facial movements}.
\newblock \bibinfo{journal}{\emph{Proceedings of the ACM on Interactive, Mobile, Wearable and Ubiquitous Technologies}} \bibinfo{volume}{6}, \bibinfo{number}{2} (\bibinfo{year}{2022}), \bibinfo{pages}{1--24}.
\newblock


\bibitem[Li et~al\mbox{.}(2018)]%
        {li2018skingest}
\bibfield{author}{\bibinfo{person}{Ling Li}, \bibinfo{person}{Shuo Jiang}, \bibinfo{person}{Peter~B Shull}, {and} \bibinfo{person}{Guoying Gu}.} \bibinfo{year}{2018}\natexlab{}.
\newblock \showarticletitle{SkinGest: artificial skin for gesture recognition via filmy stretchable strain sensors}.
\newblock \bibinfo{journal}{\emph{Advanced Robotics}} \bibinfo{volume}{32}, \bibinfo{number}{21} (\bibinfo{year}{2018}), \bibinfo{pages}{1112--1121}.
\newblock


\bibitem[MacKenzie and Soukoreff(2003)]%
        {mackenzie2003phrase}
\bibfield{author}{\bibinfo{person}{I~Scott MacKenzie} {and} \bibinfo{person}{R~William Soukoreff}.} \bibinfo{year}{2003}\natexlab{}.
\newblock \showarticletitle{Phrase sets for evaluating text entry techniques}. In \bibinfo{booktitle}{\emph{CHI'03 extended abstracts on Human factors in computing systems}}. \bibinfo{pages}{754--755}.
\newblock


\bibitem[Mahmud et~al\mbox{.}(2024a)]%
        {mahmud2024munchsonic}
\bibfield{author}{\bibinfo{person}{Saif Mahmud}, \bibinfo{person}{Devansh Agarwal}, \bibinfo{person}{Ashwin Ajit}, \bibinfo{person}{Qikang Liang}, \bibinfo{person}{Thalia Viranda}, \bibinfo{person}{Francois Guimbretiere}, {and} \bibinfo{person}{Cheng Zhang}.} \bibinfo{year}{2024}\natexlab{a}.
\newblock \showarticletitle{MunchSonic: Tracking Fine-grained Dietary Actions through Active Acoustic Sensing on Eyeglasses}. In \bibinfo{booktitle}{\emph{Proceedings of the 2024 ACM International Symposium on Wearable Computers}}. \bibinfo{pages}{96--103}.
\newblock


\bibitem[Mahmud et~al\mbox{.}(2023)]%
        {mahmud2023posesonic}
\bibfield{author}{\bibinfo{person}{Saif Mahmud}, \bibinfo{person}{Ke Li}, \bibinfo{person}{Guilin Hu}, \bibinfo{person}{Hao Chen}, \bibinfo{person}{Richard Jin}, \bibinfo{person}{Ruidong Zhang}, \bibinfo{person}{Fran{\c{c}}ois Guimbreti{\`e}re}, {and} \bibinfo{person}{Cheng Zhang}.} \bibinfo{year}{2023}\natexlab{}.
\newblock \showarticletitle{Posesonic: 3d upper body pose estimation through egocentric acoustic sensing on smartglasses}.
\newblock \bibinfo{journal}{\emph{Proceedings of the ACM on Interactive, Mobile, Wearable and Ubiquitous Technologies}} \bibinfo{volume}{7}, \bibinfo{number}{3} (\bibinfo{year}{2023}), \bibinfo{pages}{1--28}.
\newblock


\bibitem[Mahmud et~al\mbox{.}(2024b)]%
        {mahmud2024wristsonic}
\bibfield{author}{\bibinfo{person}{Saif Mahmud}, \bibinfo{person}{Kian Mahmoodi}, \bibinfo{person}{Chi-Jung Lee}, \bibinfo{person}{Francois Guimbretiere}, {and} \bibinfo{person}{Cheng Zhang}.} \bibinfo{year}{2024}\natexlab{b}.
\newblock \showarticletitle{WristSonic: Enabling Fine-grained Hand-Face Interactions on Smartwatches Using Active Acoustic Sensing}.
\newblock \bibinfo{journal}{\emph{arXiv preprint arXiv:2411.08217}} (\bibinfo{year}{2024}).
\newblock


\bibitem[Mahmud et~al\mbox{.}(2024c)]%
        {mahmud2024actsonic}
\bibfield{author}{\bibinfo{person}{Saif Mahmud}, \bibinfo{person}{Vineet Parikh}, \bibinfo{person}{Qikang Liang}, \bibinfo{person}{Ke Li}, \bibinfo{person}{Ruidong Zhang}, \bibinfo{person}{Ashwin Ajit}, \bibinfo{person}{Vipin Gunda}, \bibinfo{person}{Devansh Agarwal}, \bibinfo{person}{Fran{\c{c}}ois Guimbreti{\`e}re}, {and} \bibinfo{person}{Cheng Zhang}.} \bibinfo{year}{2024}\natexlab{c}.
\newblock \showarticletitle{ActSonic: Recognizing Everyday Activities from Inaudible Acoustic Wave Around the Body}.
\newblock \bibinfo{journal}{\emph{Proceedings of the ACM on Interactive, Mobile, Wearable and Ubiquitous Technologies}} \bibinfo{volume}{8}, \bibinfo{number}{4} (\bibinfo{year}{2024}), \bibinfo{pages}{1--32}.
\newblock


\bibitem[Maria and Deja(2024)]%
        {maria2024alexa}
\bibfield{author}{\bibinfo{person}{Tyrone Justin~Sta Maria} {and} \bibinfo{person}{Jordan~Aiko Deja}.} \bibinfo{year}{2024}\natexlab{}.
\newblock \showarticletitle{Alexa, I Wanna See You: Envisioning Smart Home Assistants for the Deaf and Hard-of-Hearing}.
\newblock \bibinfo{journal}{\emph{arXiv preprint arXiv:2412.00514}} (\bibinfo{year}{2024}).
\newblock


\bibitem[Martin et~al\mbox{.}(2023)]%
        {martin2023fingerspeller}
\bibfield{author}{\bibinfo{person}{David Martin}, \bibinfo{person}{Zikang Leng}, \bibinfo{person}{Tan Gemicioglu}, \bibinfo{person}{Jon Womack}, \bibinfo{person}{Jocelyn Heath}, \bibinfo{person}{William~C Neubauer}, \bibinfo{person}{Hyeokhyen Kwon}, \bibinfo{person}{Thomas Ploetz}, {and} \bibinfo{person}{Thad Starner}.} \bibinfo{year}{2023}\natexlab{}.
\newblock \showarticletitle{FingerSpeller: Camera-Free Text Entry Using Smart Rings for American Sign Language Fingerspelling Recognition}. In \bibinfo{booktitle}{\emph{Proceedings of the 25th International ACM SIGACCESS Conference on Computers and Accessibility}}. \bibinfo{pages}{1--5}.
\newblock


\bibitem[Morere and Roberts(2012)]%
        {morere2012fingerspelling}
\bibfield{author}{\bibinfo{person}{Donna~A Morere} {and} \bibinfo{person}{Rachel Roberts}.} \bibinfo{year}{2012}\natexlab{}.
\newblock \showarticletitle{Fingerspelling}.
\newblock In \bibinfo{booktitle}{\emph{Assessing literacy in deaf individuals: Neurocognitive measurement and predictors}}. \bibinfo{publisher}{Springer}, \bibinfo{pages}{179--189}.
\newblock


\bibitem[Mummadi et~al\mbox{.}(2017)]%
        {mummadi2017real}
\bibfield{author}{\bibinfo{person}{Chaithanya~Kumar Mummadi}, \bibinfo{person}{Frederic Philips~Peter Leo}, \bibinfo{person}{Keshav~Deep Verma}, \bibinfo{person}{Shivaji Kasireddy}, \bibinfo{person}{Philipp~Marcel Scholl}, {and} \bibinfo{person}{Kristof Van~Laerhoven}.} \bibinfo{year}{2017}\natexlab{}.
\newblock \showarticletitle{Real-time embedded recognition of sign language alphabet fingerspelling in an IMU-based glove}. In \bibinfo{booktitle}{\emph{Proceedings of the 4th international Workshop on Sensor-based Activity Recognition and Interaction}}. \bibinfo{pages}{1--6}.
\newblock


\bibitem[Oz and Leu(2005)]%
        {oz2005recognition}
\bibfield{author}{\bibinfo{person}{Cemil Oz} {and} \bibinfo{person}{Ming~C Leu}.} \bibinfo{year}{2005}\natexlab{}.
\newblock \showarticletitle{Recognition of finger spelling of American sign language with artificial neural network using position/orientation sensors and data glove}. In \bibinfo{booktitle}{\emph{International Symposium on Neural Networks}}. Springer, \bibinfo{pages}{157--164}.
\newblock


\bibitem[Padden and Gunsauls(2003)]%
        {padden2003alphabet}
\bibfield{author}{\bibinfo{person}{Carol~A Padden} {and} \bibinfo{person}{Darline~Clark Gunsauls}.} \bibinfo{year}{2003}\natexlab{}.
\newblock \showarticletitle{How the alphabet came to be used in a sign language}.
\newblock \bibinfo{journal}{\emph{Sign Language Studies}} (\bibinfo{year}{2003}), \bibinfo{pages}{10--33}.
\newblock


\bibitem[Pannattee et~al\mbox{.}(2021)]%
        {pannattee2021novel}
\bibfield{author}{\bibinfo{person}{Peerawat Pannattee}, \bibinfo{person}{Wuttipong Kumwilaisak}, \bibinfo{person}{Chatchawarn Hansakunbuntheung}, {and} \bibinfo{person}{Nattanun Thatphithakkul}.} \bibinfo{year}{2021}\natexlab{}.
\newblock \showarticletitle{Novel american sign language fingerspelling recognition in the wild with weakly supervised learning and feature embedding}. In \bibinfo{booktitle}{\emph{2021 18th International Conference on Electrical Engineering/Electronics, Computer, Telecommunications and Information Technology (ECTI-CON)}}. IEEE, \bibinfo{pages}{291--294}.
\newblock


\bibitem[Parikh et~al\mbox{.}(2024)]%
        {parikh2024echoguide}
\bibfield{author}{\bibinfo{person}{Vineet Parikh}, \bibinfo{person}{Saif Mahmud}, \bibinfo{person}{Devansh Agarwal}, \bibinfo{person}{Ke Li}, \bibinfo{person}{Fran{\c{c}}ois Guimbreti{\`e}re}, {and} \bibinfo{person}{Cheng Zhang}.} \bibinfo{year}{2024}\natexlab{}.
\newblock \showarticletitle{EchoGuide: Active Acoustic Guidance for LLM-Based Eating Event Analysis from Egocentric Videos}. In \bibinfo{booktitle}{\emph{Proceedings of the 2024 ACM International Symposium on Wearable Computers}}. \bibinfo{pages}{40--47}.
\newblock


\bibitem[Paudyal et~al\mbox{.}(2017)]%
        {paudyal2017dyfav}
\bibfield{author}{\bibinfo{person}{Prajwal Paudyal}, \bibinfo{person}{Junghyo Lee}, \bibinfo{person}{Ayan Banerjee}, {and} \bibinfo{person}{Sandeep~KS Gupta}.} \bibinfo{year}{2017}\natexlab{}.
\newblock \showarticletitle{Dyfav: Dynamic feature selection and voting for real-time recognition of fingerspelled alphabet using wearables}. In \bibinfo{booktitle}{\emph{Proceedings of the 22nd International Conference on Intelligent User Interfaces}}. \bibinfo{pages}{457--467}.
\newblock


\bibitem[Quinto-Pozos(2010)]%
        {quinto2010rates}
\bibfield{author}{\bibinfo{person}{David Quinto-Pozos}.} \bibinfo{year}{2010}\natexlab{}.
\newblock \showarticletitle{Rates of fingerspelling in american sign language}. In \bibinfo{booktitle}{\emph{Poster presented at 10th Theoretical Issues in Sign Language Research conference, West Lafayette, Indiana}}, Vol.~\bibinfo{volume}{30}.
\newblock


\bibitem[Rinalduzzi et~al\mbox{.}(2021)]%
        {rinalduzzi2021gesture}
\bibfield{author}{\bibinfo{person}{Matteo Rinalduzzi}, \bibinfo{person}{Alessio De~Angelis}, \bibinfo{person}{Francesco Santoni}, \bibinfo{person}{Emanuele Buchicchio}, \bibinfo{person}{Antonio Moschitta}, \bibinfo{person}{Paolo Carbone}, \bibinfo{person}{Paolo Bellitti}, {and} \bibinfo{person}{Mauro Serpelloni}.} \bibinfo{year}{2021}\natexlab{}.
\newblock \showarticletitle{Gesture recognition of sign language alphabet using a magnetic positioning system}.
\newblock \bibinfo{journal}{\emph{Applied Sciences}} \bibinfo{volume}{11}, \bibinfo{number}{12} (\bibinfo{year}{2021}), \bibinfo{pages}{5594}.
\newblock


\bibitem[Rizwan et~al\mbox{.}(2019)]%
        {rizwan2019american}
\bibfield{author}{\bibinfo{person}{Shaheer~Bin Rizwan}, \bibinfo{person}{Muhammad Saad~Zahid Khan}, {and} \bibinfo{person}{Muhammad Imran}.} \bibinfo{year}{2019}\natexlab{}.
\newblock \showarticletitle{American sign language translation via smart wearable glove technology}. In \bibinfo{booktitle}{\emph{2019 International Symposium on Recent Advances in Electrical Engineering (RAEE)}}, Vol.~\bibinfo{volume}{4}. IEEE, \bibinfo{pages}{1--6}.
\newblock


\bibitem[Saggio et~al\mbox{.}(2020)]%
        {saggio2020sign}
\bibfield{author}{\bibinfo{person}{Giovanni Saggio}, \bibinfo{person}{Pietro Cavallo}, \bibinfo{person}{Mariachiara Ricci}, \bibinfo{person}{Vito Errico}, \bibinfo{person}{Jonathan Zea}, {and} \bibinfo{person}{Marco~E Benalc{\'a}zar}.} \bibinfo{year}{2020}\natexlab{}.
\newblock \showarticletitle{Sign language recognition using wearable electronics: implementing k-nearest neighbors with dynamic time warping and convolutional neural network algorithms}.
\newblock \bibinfo{journal}{\emph{Sensors}} \bibinfo{volume}{20}, \bibinfo{number}{14} (\bibinfo{year}{2020}), \bibinfo{pages}{3879}.
\newblock


\bibitem[Saquib and Rahman(2020)]%
        {saquib2020application}
\bibfield{author}{\bibinfo{person}{Nazmus Saquib} {and} \bibinfo{person}{Ashikur Rahman}.} \bibinfo{year}{2020}\natexlab{}.
\newblock \showarticletitle{Application of machine learning techniques for real-time sign language detection using wearable sensors}. In \bibinfo{booktitle}{\emph{Proceedings of the 11th ACM Multimedia Systems Conference}}. \bibinfo{pages}{178--189}.
\newblock


\bibitem[Savur and Sahin(2016)]%
        {savur2016american}
\bibfield{author}{\bibinfo{person}{Celal Savur} {and} \bibinfo{person}{Ferat Sahin}.} \bibinfo{year}{2016}\natexlab{}.
\newblock \showarticletitle{American Sign Language Recognition system by using surface EMG signal}. In \bibinfo{booktitle}{\emph{2016 IEEE International Conference on Systems, Man, and Cybernetics (SMC)}}. IEEE, \bibinfo{pages}{002872--002877}.
\newblock


\bibitem[Shi et~al\mbox{.}(2018)]%
        {shi2018american}
\bibfield{author}{\bibinfo{person}{Bowen Shi}, \bibinfo{person}{Aurora~Martinez Del~Rio}, \bibinfo{person}{Jonathan Keane}, \bibinfo{person}{Jonathan Michaux}, \bibinfo{person}{Diane Brentari}, \bibinfo{person}{Greg Shakhnarovich}, {and} \bibinfo{person}{Karen Livescu}.} \bibinfo{year}{2018}\natexlab{}.
\newblock \showarticletitle{American sign language fingerspelling recognition in the wild}. In \bibinfo{booktitle}{\emph{2018 IEEE Spoken Language Technology Workshop (SLT)}}. IEEE, \bibinfo{pages}{145--152}.
\newblock


\bibitem[Shi et~al\mbox{.}(2019)]%
        {shi2019fingerspelling}
\bibfield{author}{\bibinfo{person}{Bowen Shi}, \bibinfo{person}{Aurora Martinez~Del Rio}, \bibinfo{person}{Jonathan Keane}, \bibinfo{person}{Diane Brentari}, \bibinfo{person}{Greg Shakhnarovich}, {and} \bibinfo{person}{Karen Livescu}.} \bibinfo{year}{2019}\natexlab{}.
\newblock \showarticletitle{Fingerspelling recognition in the wild with iterative visual attention}. In \bibinfo{booktitle}{\emph{Proceedings of the IEEE/CVF International Conference on Computer Vision}}. \bibinfo{pages}{5400--5409}.
\newblock


\bibitem[Singh and Chaturvedi(2023)]%
        {singh2023reliable}
\bibfield{author}{\bibinfo{person}{Shashank~Kumar Singh} {and} \bibinfo{person}{Amrita Chaturvedi}.} \bibinfo{year}{2023}\natexlab{}.
\newblock \showarticletitle{A reliable and efficient machine learning pipeline for american sign language gesture recognition using EMG sensors}.
\newblock \bibinfo{journal}{\emph{Multimedia Tools and Applications}} \bibinfo{volume}{82}, \bibinfo{number}{15} (\bibinfo{year}{2023}), \bibinfo{pages}{23833--23871}.
\newblock


\bibitem[Sun et~al\mbox{.}(2023)]%
        {sun2023echonose}
\bibfield{author}{\bibinfo{person}{Rujia Sun}, \bibinfo{person}{Xiaohe Zhou}, \bibinfo{person}{Benjamin Steeper}, \bibinfo{person}{Ruidong Zhang}, \bibinfo{person}{Sicheng Yin}, \bibinfo{person}{Ke Li}, \bibinfo{person}{Shengzhang Wu}, \bibinfo{person}{Sam Tilsen}, \bibinfo{person}{Francois Guimbretiere}, {and} \bibinfo{person}{Cheng Zhang}.} \bibinfo{year}{2023}\natexlab{}.
\newblock \showarticletitle{EchoNose: Sensing Mouth, Breathing and Tongue Gestures inside Oral Cavity using a Non-contact Nose Interface}. In \bibinfo{booktitle}{\emph{Proceedings of the 2023 ACM International Symposium on Wearable Computers}}. \bibinfo{pages}{22--26}.
\newblock


\bibitem[Taylor(2018)]%
        {taylor2018real}
\bibfield{author}{\bibinfo{person}{Brandon Taylor}.} \bibinfo{year}{2018}\natexlab{}.
\newblock \emph{\bibinfo{title}{Real-Time Depth-Based Hand Tracking for American Sign Language Recognition}}.
\newblock \bibinfo{thesistype}{Ph.\,D. Dissertation}. \bibinfo{school}{Carnegie Mellon University}.
\newblock


\bibitem[Valli et~al\mbox{.}(1992)]%
        {valli1992linguistics}
\bibfield{author}{\bibinfo{person}{Clayton Valli}, \bibinfo{person}{Ceil Lucas}, {and} \bibinfo{person}{Valerie Dively}.} \bibinfo{year}{1992}\natexlab{}.
\newblock \bibinfo{booktitle}{\emph{Linguistics of American sign language}}.
\newblock \bibinfo{publisher}{Gallaudet University Press Washington, DC}.
\newblock


\bibitem[Wang et~al\mbox{.}(2018)]%
        {wang2018c}
\bibfield{author}{\bibinfo{person}{Tianben Wang}, \bibinfo{person}{Daqing Zhang}, \bibinfo{person}{Yuanqing Zheng}, \bibinfo{person}{Tao Gu}, \bibinfo{person}{Xingshe Zhou}, {and} \bibinfo{person}{Bernadette Dorizzi}.} \bibinfo{year}{2018}\natexlab{}.
\newblock \showarticletitle{C-FMCW based contactless respiration detection using acoustic signal}.
\newblock \bibinfo{journal}{\emph{Proceedings of the ACM on Interactive, Mobile, Wearable and Ubiquitous Technologies}} \bibinfo{volume}{1}, \bibinfo{number}{4} (\bibinfo{year}{2018}), \bibinfo{pages}{1--20}.
\newblock


\bibitem[Yoon et~al\mbox{.}(2012)]%
        {yoon2012adaptive}
\bibfield{author}{\bibinfo{person}{Jong-Won Yoon}, \bibinfo{person}{Sung-Ihk Yang}, {and} \bibinfo{person}{Sung-Bae Cho}.} \bibinfo{year}{2012}\natexlab{}.
\newblock \showarticletitle{Adaptive mixture-of-experts models for data glove interface with multiple users}.
\newblock \bibinfo{journal}{\emph{Expert Systems with Applications}} \bibinfo{volume}{39}, \bibinfo{number}{5} (\bibinfo{year}{2012}), \bibinfo{pages}{4898--4907}.
\newblock


\bibitem[Yu et~al\mbox{.}(2024)]%
        {yu2024ring}
\bibfield{author}{\bibinfo{person}{Tianhong~Catherine Yu}, \bibinfo{person}{Guilin Hu}, \bibinfo{person}{Ruidong Zhang}, \bibinfo{person}{Hyunchul Lim}, \bibinfo{person}{Saif Mahmud}, \bibinfo{person}{Chi-Jung Lee}, \bibinfo{person}{Ke Li}, \bibinfo{person}{Devansh Agarwal}, \bibinfo{person}{Shuyang Nie}, \bibinfo{person}{Jinseok Oh}, {et~al\mbox{.}}} \bibinfo{year}{2024}\natexlab{}.
\newblock \showarticletitle{Ring-a-Pose: A Ring for Continuous Hand Pose Tracking}.
\newblock \bibinfo{journal}{\emph{arXiv preprint arXiv:2404.12980}} (\bibinfo{year}{2024}).
\newblock


\bibitem[Zhang et~al\mbox{.}(2017a)]%
        {zhang2017fingersound}
\bibfield{author}{\bibinfo{person}{Cheng Zhang}, \bibinfo{person}{Anandghan Waghmare}, \bibinfo{person}{Pranav Kundra}, \bibinfo{person}{Yiming Pu}, \bibinfo{person}{Scott Gilliland}, \bibinfo{person}{Thomas Ploetz}, \bibinfo{person}{Thad~E Starner}, \bibinfo{person}{Omer~T Inan}, {and} \bibinfo{person}{Gregory~D Abowd}.} \bibinfo{year}{2017}\natexlab{a}.
\newblock \showarticletitle{FingerSound: Recognizing unistroke thumb gestures using a ring}.
\newblock \bibinfo{journal}{\emph{Proceedings of the ACM on Interactive, Mobile, Wearable and Ubiquitous Technologies}} \bibinfo{volume}{1}, \bibinfo{number}{3} (\bibinfo{year}{2017}), \bibinfo{pages}{1--19}.
\newblock


\bibitem[Zhang et~al\mbox{.}(2017b)]%
        {zhang2017fingorbits}
\bibfield{author}{\bibinfo{person}{Cheng Zhang}, \bibinfo{person}{Xiaoxuan Wang}, \bibinfo{person}{Anandghan Waghmare}, \bibinfo{person}{Sumeet Jain}, \bibinfo{person}{Thomas Ploetz}, \bibinfo{person}{Omer~T Inan}, \bibinfo{person}{Thad~E Starner}, {and} \bibinfo{person}{Gregory~D Abowd}.} \bibinfo{year}{2017}\natexlab{b}.
\newblock \showarticletitle{FingOrbits: interaction with wearables using synchronized thumb movements}. In \bibinfo{booktitle}{\emph{Proceedings of the 2017 ACM International Symposium on Wearable Computers}}. \bibinfo{pages}{62--65}.
\newblock


\bibitem[Zhang et~al\mbox{.}(2018)]%
        {zhang2018fingerping}
\bibfield{author}{\bibinfo{person}{Cheng Zhang}, \bibinfo{person}{Qiuyue Xue}, \bibinfo{person}{Anandghan Waghmare}, \bibinfo{person}{Ruichen Meng}, \bibinfo{person}{Sumeet Jain}, \bibinfo{person}{Yizeng Han}, \bibinfo{person}{Xinyu Li}, \bibinfo{person}{Kenneth Cunefare}, \bibinfo{person}{Thomas Ploetz}, \bibinfo{person}{Thad Starner}, {et~al\mbox{.}}} \bibinfo{year}{2018}\natexlab{}.
\newblock \showarticletitle{FingerPing: Recognizing fine-grained hand poses using active acoustic on-body sensing}. In \bibinfo{booktitle}{\emph{Proceedings of the 2018 CHI conference on human factors in computing systems}}. \bibinfo{pages}{1--10}.
\newblock


\bibitem[Zhang et~al\mbox{.}(2019)]%
        {zhang2019myosign}
\bibfield{author}{\bibinfo{person}{Qian Zhang}, \bibinfo{person}{Dong Wang}, \bibinfo{person}{Run Zhao}, {and} \bibinfo{person}{Yinggang Yu}.} \bibinfo{year}{2019}\natexlab{}.
\newblock \showarticletitle{MyoSign: enabling end-to-end sign language recognition with wearables}. In \bibinfo{booktitle}{\emph{Proceedings of the 24th international conference on intelligent user interfaces}}. \bibinfo{pages}{650--660}.
\newblock


\bibitem[Zhang et~al\mbox{.}(2023a)]%
        {zhang2023hpspeech}
\bibfield{author}{\bibinfo{person}{Ruidong Zhang}, \bibinfo{person}{Hao Chen}, \bibinfo{person}{Devansh Agarwal}, \bibinfo{person}{Richard Jin}, \bibinfo{person}{Ke Li}, \bibinfo{person}{Fran{\c{c}}ois Guimbreti{\`e}re}, {and} \bibinfo{person}{Cheng Zhang}.} \bibinfo{year}{2023}\natexlab{a}.
\newblock \showarticletitle{HPSpeech: Silent Speech Interface for Commodity Headphones}. In \bibinfo{booktitle}{\emph{Proceedings of the 2023 ACM International Symposium on Wearable Computers}}. \bibinfo{pages}{60--65}.
\newblock


\bibitem[Zhang et~al\mbox{.}(2023b)]%
        {zhang2023echospeech}
\bibfield{author}{\bibinfo{person}{Ruidong Zhang}, \bibinfo{person}{Ke Li}, \bibinfo{person}{Yihong Hao}, \bibinfo{person}{Yufan Wang}, \bibinfo{person}{Zhengnan Lai}, \bibinfo{person}{Fran{\c{c}}ois Guimbreti{\`e}re}, {and} \bibinfo{person}{Cheng Zhang}.} \bibinfo{year}{2023}\natexlab{b}.
\newblock \showarticletitle{EchoSpeech: Continuous Silent Speech Recognition on Minimally-Obtrusive Eyewear Powered by Acoustic Sensing}. In \bibinfo{booktitle}{\emph{Proceedings of the 2023 CHI Conference on Human Factors in Computing Systems}}. \bibinfo{pages}{1--18}.
\newblock


\bibitem[Zhou et~al\mbox{.}(2023)]%
        {zhou2023signquery}
\bibfield{author}{\bibinfo{person}{Hao Zhou}, \bibinfo{person}{Taiting Lu}, \bibinfo{person}{Kristina Mckinnie}, \bibinfo{person}{Joseph Palagano}, \bibinfo{person}{Kenneth Dehaan}, {and} \bibinfo{person}{Mahanth Gowda}.} \bibinfo{year}{2023}\natexlab{}.
\newblock \showarticletitle{SignQuery: A Natural User Interface and Search Engine for Sign Languages with Wearable Sensors}. In \bibinfo{booktitle}{\emph{Proceedings of the 29th Annual International Conference on Mobile Computing and Networking}}. \bibinfo{pages}{1--16}.
\newblock


\bibitem[Zhu et~al\mbox{.}(2018)]%
        {zhu2018typing}
\bibfield{author}{\bibinfo{person}{Suwen Zhu}, \bibinfo{person}{Tianyao Luo}, \bibinfo{person}{Xiaojun Bi}, {and} \bibinfo{person}{Shumin Zhai}.} \bibinfo{year}{2018}\natexlab{}.
\newblock \showarticletitle{Typing on an invisible keyboard}. In \bibinfo{booktitle}{\emph{Proceedings of the 2018 CHI Conference on Human Factors in Computing Systems}}. \bibinfo{pages}{1--13}.
\newblock


\end{thebibliography}
\end{document}